\documentclass[aip,jcp]{revtex4-1}
\usepackage{amsmath}
\usepackage{amsfonts}
\usepackage{amssymb}
\usepackage{graphicx}
\usepackage{color}
\usepackage{hyperref}

\newcommand \op[1] {\hat{#1}}
\newcommand \ket[1] {|{#1}\rangle}
\newcommand \bra[1] {\langle {#1}|}

\newcommand \trace[1] {\mathrm{Tr}({#1})}

\begin{document}

\title{Non-perturbative calculation of orbital- and spin effects in molecules subject to non-uniform magnetic fields}

\author{Sangita Sen}

\email{sangita.sen310187@gmail.com}
\affiliation{
Hylleraas Centre for Quantum Molecular Sciences, Department of Chemistry, University of Oslo, P.O.~Box 1033 Blindern, N-0315 Oslo, Norway}

\author{Erik I. Tellgren}

\email{erik.tellgren@kjemi.uio.no}
\affiliation{
Hylleraas Centre for Quantum Molecular Sciences, Department of Chemistry, University of Oslo, P.O.~Box 1033 Blindern, N-0315 Oslo, Norway}

\begin{abstract}
External non-uniform magnetic fields acting on molecules induce non-collinear spin-densities and spin-symmetry breaking. This necessitates a general two-component Pauli spinor representation. In this paper, we report the implementation of a General Hartree-Fock method, without any spin constraints, for non-perturbative calculations with finite non-uniform fields. London atomic orbitals are used to ensure faster basis convergence as well as invariance under constant gauge shifts of the magnetic vector potential. The implementation has been applied to an investigate the joint orbital and spin response to a field gradient---quantified through the anapole moments---of a set of small molecules placed in a linearly varying magnetic field.
The relative contributions of orbital and spin-Zeeman interaction terms have been studied both theoretically and computationally. Spin effects are stronger and show a general paramagnetic behaviour for closed shell molecules while orbital effects can have either direction. Basis set convergence and size effects of anapole susceptibility tensors have been reported. The relation of the mixed anapole susceptibility tensor to chirality is also demonstrated.
\end{abstract}

\maketitle 

\section{Introduction} \label{intro}

Quantum-chemical calculations of magnetic field-effects on molecules routinely rely on the assumptions that the magnetic field is either uniform or a dipole field, and weak enough to be treated by low-order perturbation theory. As magnetic fields accessible with present laboratory technology are weak compared to a molecular energy scale, this approach is reasonable for many purposes. However, effects beyond the reach of these idealizations are largely unexplored. An example is the behavior of atoms and molecules subject to strong uniform magnetic fields~\cite{LAI_RMP73_629} and another is the response of molecular systems subject to magnetic field gradients~\cite{SANDRATSKII_JPCM4_6927}. The response of the electrons to inhomogenities in the external magnetic field may be broadly divided into orbital effects and spin effects. Both of these effects are relatively little explored. The main exception is the work by Lazzeretti and co-workers on a perturbative formalism for the orbital response due to field gradients~\cite{Lazzeretti1989,Lazzeretti1993}. Some computational studies have also appeared at the H\"uckel-level~\cite{Ceulemans1998}, Hartree--Fock level~\cite{Faglioni2004,Caputo1994,Caputo1994a,Caputo1996,Caputo1997}, and correlated levels~\cite{ZARYCZ_JCC37_1552}.

Spin effects have not been explored directly in quantum-chemical studies, having so far been the domain of solid-state physics~\cite{SANDRATSKII_JPCM4_6927}. The present works aims to take the initial step to filling this gap by studying combined orbital and spin effects in a few molecular systems. To capture spin effects, standard simplifying constraints on the spin degrees of freedom need to be dropped in favor of general two-component Pauli spinors. Magnetic field gradients induce non-collinear spin densities, with the spin Zeeman interaction driving (anti-)alignment towards the local magnetic field direction, as well as spin-symmetry breaking. Neither the magnitude of the total spin nor the spin projection onto an axis are good quantum numbers in the presence of magnetic field gradients. Geometric frustration and static correlation can also result in non-collinear spin densities~\cite{YAMAGUCHI_CPL49_555,YAMAKI_IJQC80_701,YAMAKI_IJQC84_546,JIMENEZHOYOS_JCTC7_2667,LUO_JCTC9_5349}, though in this case the spin operator $\op{S}^2$ commutes with the hamiltonian and any spin-symmetry breaking is an artifact of limitations of the Hartree--Fock and most post-Hartree--Fock methods.

The perturbative approach quickly becomes unwieldy for higher-order magnetic response, in particular when London atomic orbitals (LAOs) are employed to enforce gauge-origin invariance and accelerate basis set convergence~\cite{London1937,Hameka1958,Ditchfield1976,Helgaker1991}. When ordinary Gaussians are used, very large basis sets become necessary to approach  gauge-origin invariance~\cite{Faglioni2004,Caputo1994,Caputo1994a,Caputo1996,Caputo1997}. We  use LAOs in combination with a non-perturbative (finite field) approach. This necessitates an implementation of integral evaluation for the LAOs which are plane-wave/Gaussian hybrid functions~\cite{Tellgren2008,REYNOLDS_PCCP17_14280}, but requires no additional modification for higher order properties.
An added implementation advantage is that only the one-electron part of the Hamiltonian is modified for various external fields and no additional effort is required for extension to post-Hartree--Fock theories.
It therefore opens up the possibility of studying non-perturbative phenomena.
The finite field procedure for both uniform and non-uniform magnetic fields involving LAOs have been developed recently~\cite{Tellgren2008,Tellgren2012}. 
This has led to the discovery of non-perturbative transition from closed-shell para- to diamagnetism~\cite{Tellgren2009} and a new bonding mechanism~\cite{Lange2012,Tellgren2012,Kubo2007} in very strong magnetic fields.

A convenient quantification of the response to (transverse) magnetic field gradients is provided by the {\it anapole moment}. In a multipole expansion of the energy, anapole moments are those moments which couple linearly to the curl of the magnetic field~\cite{Gray2010}.
They have been largely neglected ever since they were first considered by Zeldovich  in 1957~\cite{Zeldovich1957}, who also introduced the term `anapole'. 
These moments are distinct from the usual magnetic moments arising out of a perturbative expansion in the magnetic field and can be visualised instead as arising from the meridional currents in a toroidal charge distribution.
They are anti-symmetric under both spatial inversion and time-reversal.
Due to the relation of nuclear anapole moments with parity violation in atoms and molecules, they have received attention from nuclear physicists~\cite{Haxton1997,Haxton2002} with the first experimental evidence coming from measurements on the Cs atom~\cite{Wood1997,Haxton2001,DeMille2008}.
Spaldin et al.\ have suggested experiments for measuring anapole moments in ferrotoroidic materials~\cite{Spaldin2008}.
Experiments for measuring permanent and induced electronic anapole moments have also been suggested\cite{Pelloni2011,Khriplovich1990}.
Only special structures have permanent anapole moments such as molecular nanotoroids~\cite{Ceulemans1998,Berger2012}, ferroelectric nanostructures~\cite{Naumov2004,VanAken2007}, ferromagnetic structures~\cite{Klui2003} and Dy clusters (single molecule magnets)~\cite{Novitchi2012,Guo2012,Ungur2012}.
Several groups have experimentally demonstrated anapole moments in metamaterials for potential application in sensors~\cite{Kaelberer2010,Ye2013,Basharin2015a}.
In an external non-uniform field, toroidal spin and/or orbital currents are induced, giving rise to anapole moments and the corresponding susceptibilities can be computed.
The induced anapolar current densities have been studied for conjugated cyclic acetylenes~\cite{Berger2012}. Another study analyzed topological features of anapolar current densities in small molecules~\cite{Pelloni2011}.
Spin and orbital contributions to anapole moments have been analyzed in a simple analytical model of diatomics\cite{Khriplovich1990,Lewis1994}.
Ab initio computational studies, both perturbative approaches~\cite{Faglioni2004,Caputo1994,Caputo1994a,Caputo1996,Caputo1997} and non-perturbative approaches~\cite{Tellgren2013}, have estimated the orbital contribution to anapole susceptibilities in closed-shell molecules.
Some recent efforts have also been aimed at further understanding the interactions of toroidal moments with external fields~\cite{Marinov2007,Spaldin2008,Kaelberer2010,Ogut2012,Ye2013}.

In this work we study the combined the orbital and spin effects in a set of molecules subject to magnetic field gradients. We focus on transverse gradients and induced anapole moments. The relative importance of orbital and spin effects is investigated numerically. We also provide theoretical results on the direction and additivity of the spin contributions to induced anapole moments. The out line of the article is as follows.
In Sec.~\ref{definitions}, we define the Hamiltonian and the properties relevant to our study.
The General Hartree-Fock model, necessary for mean-field computations in non-uniform magnetic fields, is described in Sec.~\ref{GHFtheory}.
Secs.~\ref{SpinEffectsOnA} and \ref{SpinandOrb} discuss the theory behind the spin and orbital interactions with the external magnetic field and their relative effect on the molecular energy and properties.
In Sec.~\ref{results} we present our numerical results and, finally, we conclude with the summary in Sec.~\ref{summary}.

\section{Hamiltonian and Properties} \label{definitions}

In what follows, we shall consider electronic systems subject to a linearly varying magnetic field of the form
\begin{equation}
  \mathbf{B}_{\text{tot}}(\mathbf{r}) = \mathbf{B} + \mathbf{r}_{\mathbf{h}}^T \mathbf{b} - \frac{1}{3}\mathbf{r}_{\mathbf{h}} \, \trace{\mathbf{b}},
\end{equation}
where $\mathbf{B}$ is a uniform (position independent) component, $\mathbf{b}$ is a $3\times 3$ matrix defining the field gradients, and $\mathbf{r}_{\mathbf{h}} = \mathbf{r} - \mathbf{h}$ is the position relative to some reference point $\mathbf{h}$. This type of magnetic field is most naturally viewed as arising from a Taylor expansion around $\mathbf{r} = \mathbf{h}$, truncated at linear order. The corresponding vector potential can be written as
\begin{align}
\mathbf{A}_{\text{tot}}(\mathbf{r}) &= \frac{1}{2}\mathbf{B} \times \mathbf{r_g}-\frac{1}{3}\mathbf{r_h} \times (\mathbf{r_h}^T\mathbf{b}),
\end{align}
where $\mathbf{r}_{\mathbf{g}} = \mathbf{r} - \mathbf{g}$, $\mathbf{g}$ being the gauge origin. It can be verified that $\mathbf{B}_{\text{tot}} = \nabla\times\mathbf{A}_{\text{tot}}$ and that the magnetic field is divergence free, $\nabla\cdot\mathbf{B}_{\text{tot}} = 0$. In what follows, we focus on the anti-symmetric part C$_{\alpha} = \epsilon_{\alpha\beta\gamma} b_{\beta\gamma}$ of the matrix $\mathbf{b}$ and take the symmetric part, $\mathbf{b} = \mathbf{b}^T$, to vanish. We can then write
\begin{align}
  \mathbf{A}_{\text{tot}}(\mathbf{r}) & = \frac{1}{2}\mathbf{B} \times \mathbf{r}_{\mathbf{g}}-\frac{1}{3} \mathbf{r}_{\mathbf{h}} \times (\mathbf{C}\times\mathbf{r}_{\mathbf{h}}),
         \\
  \mathbf{B}_{\text{tot}}(\mathbf{r}) & = \mathbf{B} + \frac{1}{2} \mathbf{C}\times\mathbf{r}_{\mathbf{h}}.
\end{align}
Furthermore, the constant vector encoding the anti-symmetric part of $\mathbf{b}$  equals the curl of the magnetic field, $\nabla \times \mathbf{B}_{\text{tot}} = \mathbf{C}$.

The non-relativistic Schr\"odinger--Pauli Hamiltonian is given by
\begin{equation}
\hat{H} = \frac{1}{2}\sum_l \hat{\pi_l}^2 - \sum_l v(\mathbf{r}_l) + \sum_{k<l} \frac{1}{r_{kl}} + \sum_l\mathbf{B}_{\text{tot}}(\mathbf{r}_l)\cdot \hat{\mathbf{S}}_l \label{Hamiltonian}
\end{equation}
where $\hat{\boldsymbol{\pi}}_l = -i\nabla_l + \mathbf{A}_{\text{tot}}(\mathbf{r}_l)$ is the mechanical momentum operator. Properties can be alternately viewed as expectation values $\bra{\Psi} \hat{\Omega} \ket{\Psi}$ or as derivatives of the energy $E = \bra{\Psi} \hat{H} \ket{\Psi}$ related to terms in a Taylor expansion. The first order properties are given by the orbital and spin contributions to the magnetic dipole moment,
\begin{align}
   \mathbf{L}_{\mathbf{q}} & = \sum_l \bra{\Psi} \hat{\mathbf{L}}_{\mathbf{q};l}\ket{\Psi}, \quad \hat{\mathbf{L}}_{\mathbf{q};l} = \mathbf{r}_{\mathbf{q};l} \times \hat{\boldsymbol{\pi}}_l,
        \\
  \mathbf{S} & = \sum_l \bra{\Psi} \hat{\mathbf{S}}_l \ket{\Psi},
\end{align}
which combine to a total dipole moment $\mathbf{J}_{\mathbf{q}} = \mathbf{L}_{\mathbf{q}} + 2 \mathbf{S}$. Here, $\mathbf{q}$ is an arbitrary reference point. Given the form of the magnetic vector potential above, it is $\mathbf{L}_{\mathbf{g}}$, with the reference point at the gauge origin, that is the relevant magnetic dipole moment. The anapole moment is similarly given by,
\begin{equation}
   \mathbf{a} = -\sum_l \bra{\Psi} \mathbf{r}_{\mathbf{h};l} \times (\tfrac{1}{3} \hat{\mathbf{L}}_{\mathbf{h};l} + \hat{\mathbf{S}}_l) \ket{\Psi}.
\end{equation}
The weighting of the orbital and spin contributions to these quantities is not arbitrary and takes a more intuitive form when they are expressed in terms of the total current density,
\begin{equation}
   \mathbf{j}(\mathbf{r}) = \frac{\delta E}{\delta \mathbf{A}_{\text{tot}}(\mathbf{r})} = \sum_l \frac{1}{2} \bra{\Psi} \{ \delta(\mathbf{r}_l - \mathbf{r}), \hat{\boldsymbol{\pi}}_l \} \ket{\Psi} + \nabla\times\sum_l \bra{\Psi} \delta(\mathbf{r}_l-\mathbf{r}) \hat{\mathbf{S}} \ket{\Psi},
\end{equation}
where the first term is the orbital current density and the last term---the curl of the spin density---is the spin current density. The magnetic dipole moment and anapole moment can now be identified with linear and quadratic moments of the total current density,
\begin{align}
    \mathbf{J}_{\mathbf{g}} & = \int \mathbf{r}_{\mathbf{g}} \times \mathbf{j}(\mathbf{r}) \, d\mathbf{r},
       \\
    \mathbf{a} & = -\frac{1}{3} \int \mathbf{r}_{\mathbf{h}} \times \big( \mathbf{r}_{\mathbf{h}} \times \mathbf{j}(\mathbf{r}) \big) \, d\mathbf{r}.
\end{align}

In a non-perturbative setting, the energy $E$ as well as expectation value properties like $\mathbf{J}_{\mathbf{g}}$ and $\mathbf{a}$ can be obtained directly as functions of $\mathbf{B}$ and $\mathbf{C}$. On the other hand, a Taylor expansion of the energy defines second-order properties
\begin{equation}
  \label{eqEnergyExpansion}
   E(\mathbf{B},\mathbf{C}) \approx E_0 + \mathbf{B} \cdot \mathbf{J}_{\mathbf{g}} - \frac{1}{2} \mathbf{C}\cdot \mathbf{a} -  \frac{1}{2} \mathbf{B}^T \boldsymbol{\chi} \mathbf{B} - \mathbf{B} \mathcal{M} \mathbf{C} - \frac{1}{2} \mathbf{C}^T \mathcal{A} \mathbf{C},
\end{equation}
where $\mathbf{J}_{\mathbf{g}}$ and $\mathbf{a}$ are here evaluated at vanishing $\mathbf{B}_{\text{tot}}$. Specifically, $\boldsymbol{\chi}$ is the magnetizability tensor, $\mathcal{M}$ the mixed anapole susceptibility tensor, and $\mathcal{A}$ the anapole susceptibility tensor.

When the preconditions of the Hellmann--Feynman theorem are satisfied, the above expectation value quantities can be identified with energy derivatives,
\begin{align}
   \mathbf{J}_{\mathbf{g}} & \stackrel{!}{=} 2 \frac{\partial E(\mathbf{B},\mathbf{C})}{\partial \mathbf{B}},
         \\
   \mathbf{a} & \stackrel{!}{=} -2 \frac{\partial E(\mathbf{B},\mathbf{C})}{\partial \mathbf{C}}.
\end{align}
However, when LAOs are used, the basis set depends on the parameters $\mathbf{B}$ and $\mathbf{C}$ leading to the expectation values and the energy derivatives being slightly different, in general, though the discrepancy vanishes in the limit of a complete basis.

Turning to the second-order susceptibilities, the defining expressions are
\begin{align}
		\mathcal{A}_{\alpha\beta} &= -2\frac{\partial^2 E(\mathbf{B},\mathbf{C})}{\partial C_\alpha\partial C_\beta}\bigg|_{\mathbf{B}=0,\mathbf{C}=0}, \label{Asus_def}\\
		\mathcal{M}_{\alpha\beta} &= \frac{\partial^2 E(\mathbf{B},\mathbf{C})}{\partial B_\alpha\partial C_\beta}\bigg|_{\mathbf{B}=0,\mathbf{C}=0}. \label{Msus_def}
\end{align}
When the preconditions of the Hellmann--Feynman theorem do not apply, one can also introduce the closely related, but inequivalent, quantities
\begin{align}
		\mathcal{A}_{\alpha\beta}' &= \frac{\partial a_\alpha(\mathbf{B},\mathbf{C})}{\partial C_\beta}\bigg|_{\mathbf{B}=0,\mathbf{C}=0}, \label{Aprimesus_def}\\
		\mathcal{M}_{\alpha\beta}' &= -\frac{\partial L_{\mathbf{g};\alpha}(\mathbf{B},\mathbf{C})}{\partial C_\beta}\bigg|_{\mathbf{B}=0,\mathbf{C}=0},  \label{Mprimesus_def}\\
		\mathcal{M}_{\alpha\beta}'' &= \frac{\partial a_{\beta}(\mathbf{B},\mathbf{C})}{\partial B_\alpha}\bigg|_{\mathbf{B}=0,\mathbf{C}=0}.  \label{Mdblprimesus_def}
\end{align}
Again, in the basis set limit, equivalence is restored, i.e., $\mathcal{A} = \mathcal{A}'$ and $\mathcal{M} = \mathcal{M}' = \mathcal{M}''$. However, for finite LAO basis sets, numerical investigation of the basis set convergence is warranted.

\section{The General Hartree-Fock Model With An External Non-Uniform Magnetic Field} \label{GHFtheory}

For a constant magnetic field $\mathbf{B}$ aligned to the z-axis, the spin operators $\hat{S}^2$ and $\hat{S}_z$ commute with the Hamiltonian and each molecular orbital can be chosen in a factorized form, where each spatial function multiplies a constant spin part. Moreover, each spin part defines a spin that is either parallel or anti-parallel to $\mathbf{B}$---the uniform field defines a global spin quantization axis.
In the presence of a non-uniform magnetic field, however, neither $\hat{S}^2$ nor any projection $\mathbf{u}\cdot\hat{\mathbf{S}}$ (on a fixed unit vector  $\mathbf{u}$) need to yield good quantum numbers. The spatial and spin parts of molecular orbitals become coupled, requiring a representation as 2-component Pauli spinors. Because the direction of the magnetic field varies with position, there is also no natural global spin quantization axis and the spin density becomes {\it non-collinear}. The present setting is thus unusual in that it concerns a non-relativistic Hamiltonian that requires a non-collinear, 2-component representation.

The Hartree-Fock (HF) models in the non-relativistic domain may be subdivided into Restricted HF (RHF), Unrestricted HF (UHF), Restricted Open-shell HF (ROHF) and General HF (GHF). The RHF model imposes singlet spin symmetry, and is therefore oblivious to the spin-Zeeman term, making it a useful analysis tool for estimating the purely orbital contribution to the total magnetic field effect. The UHF and ROHF flavors impose a global spin quantization axis and are therefore not meaningful in combination with a position-dependent spin-Zeeman term. Nonetheless, we shall below consider UHF results obtained with the spin-Zeeman term disabled, in order to isolate the purely orbital field effects in open-shell systems. In order to treat joint orbital and spin effects, the HF flavor can be no less than a complex GHF model.

In more detail, molecular orbitals $\phi_K(\mathbf{r})$ in the GHF model take the form of generic 2-component spinors,
\begin{equation}
\phi_K(\mathbf{r}) = \sum_{a} C^{a,K\uparrow} \chi_{a}(\mathbf{r}) \begin{pmatrix} 1 \\ 0 \end{pmatrix} + \sum_{a} C^{a,K\downarrow} \chi_{a}(\mathbf{r}) \begin{pmatrix} 0 \\ 1 \end{pmatrix} = \sum_{a}  \chi_{a}(\mathbf{r}) \begin{pmatrix} C^{a,K\uparrow} \\ C^{a,K\downarrow} \end{pmatrix}.
\end{equation}
$\chi_{a}$s denote spin-free basis functions. From now on $\Psi$ will denote a Slater determinant formed from such spinors. The spinors also define an associated $2\times 2$ density matrix kernel,
\begin{equation}
\begin{split}
D^{2\times 2}(\mathbf{r},\mathbf{r}') & = \sum_K^{\text{occ}} \phi_K(\mathbf{r}) \phi_K(\mathbf{r}')^{\dagger} 
  = \sum_{ab} \chi_{a}(\mathbf{r})
\begin{pmatrix}
D^{\uparrow\uparrow;ab}  & D^{\uparrow\downarrow;ab} \\
D^{\downarrow\uparrow;ab}  & D^{\downarrow\downarrow;ab}
\end{pmatrix}
\chi_{b}(\mathbf{r}')^*.
\end{split}
\end{equation}
For given basis function indices $a$ and $b$, the corresponding matrix elements are written
\begin{align}
D^{\sigma\tau;ab} & = \sum_K C^{a,K\sigma} C^{b,K\tau *}, \quad \sigma, \tau \in \{\uparrow,\downarrow\}.
\end{align}

The GHF electronic energy can be decomposed into kinetic, spin-Zeeman, electrostatic nuclear attraction, Coulomb repulsion, and exchange energy. Only the spin-Zeeman and exchange terms differ substantially from the standard RHF and UHF forms, since only these terms involve the off-diagonal spin blocks of $D^{2\times 2}$.

When evaluating the spin-Zeeman term, it is convenient to introduce density matrix-like quantities obtained by letting the Pauli spin matrices act on $D^{2\times 2}$ and tracing out the spin degrees of freedom:
\begin{align}
M_x(\mathbf{r},\mathbf{r}') & = \sum_{\tau} \bra{\tau} \sigma_x D^{2\times 2}(\mathbf{r},\mathbf{r}') \ket{\tau} = \sum_{ab} \chi_{a}(\mathbf{r}) \chi_{b}(\mathbf{r}')^* (D^{\downarrow\uparrow;ab} + D^{\uparrow\downarrow;ab}), \\
M_y(\mathbf{r},\mathbf{r}') & = \sum_{\tau} \bra{\tau} \sigma_y D^{2\times 2}(\mathbf{r},\mathbf{r}') \ket{\tau} = \sum_{ab} \chi_{a}(\mathbf{r}) \chi_{b}(\mathbf{r}')^* (-i D^{\downarrow\uparrow;ab} + i D^{\uparrow\downarrow;ab}), \\
M_z(\mathbf{r},\mathbf{r}') & = \sum_{\tau} \bra{\tau} \sigma_z D^{2\times 2}(\mathbf{r},\mathbf{r}') \ket{\tau} = \sum_{ab} \chi_{a}(\mathbf{r}) \chi_{b}(\mathbf{r}')^*
(D^{\uparrow\uparrow;ab} - D^{\downarrow\downarrow;ab}).
\end{align}
Letting $S_{ba} = \int \chi_{b}(\mathbf{r})^* \, \chi_{a}(\mathbf{r}) d\mathbf{r}$ denote an overlap integral and $\mu_{ba;\gamma} = \int r_{\gamma} \, \chi_{b}(\mathbf{r})^* \, \chi_{a}(\mathbf{r}) d\mathbf{r}$ a dipole moment integral, the spin-Zeeman term is given by
\begin{equation}
 \begin{split}
  E_Z & = \sum_l \bra{\Psi} \mathbf{B}_{\text{tot}}(\mathbf{r}_l)\cdot\hat{\mathbf{S}}_l \ket{\Psi} = \frac{1}{2} \int \mathbf{B}_{\text{tot}}(\mathbf{r}) \cdot \mathbf{M}(\mathbf{r},\mathbf{r}) d\mathbf{r}
         \\
     & = \frac{1}{2}  (\widetilde{B}_x S_{ba} + \mu_{ba;\gamma} \widetilde{b}_{\gamma x}) (D^{\downarrow\uparrow;ab} + D^{\uparrow\downarrow;ab})
         + \frac{1}{2}  (\widetilde{B}_y S_{ba} + \mu_{ba;\gamma} \widetilde{b}_{\gamma y}) (-iD^{\downarrow\uparrow;ab} + iD^{\uparrow\downarrow;ab})
         \\
  & \ \ + \frac{1}{2} (\widetilde{B}_z S_{ba} + \mu_{ba;\gamma} \widetilde{b}_{\gamma z}) (D^{\uparrow\uparrow;ab} - D^{\downarrow\downarrow;ab}),
 \end{split}
\end{equation}
with implicit summation over repeated indices and the notation $\widetilde{\mathbf{b}} = \mathbf{b} - \tfrac{1}{3} \trace{\mathbf{b}} \mathbf{I}$ and $\widetilde{\mathbf{B}} = \mathbf{B} - \mathbf{h}^T \widetilde{\mathbf{b}}$. The contribution to the Fock matrix is obtained as the derivative
\begin{equation}
   F^{\tau\sigma}_{Z;ba} = \frac{\partial E_Z}{\partial D^{\sigma\tau}_{ab}}.
\end{equation}
With the compact notation $\widetilde{B}_{\pm} = \widetilde{B}_x \pm i \widetilde{B}_y$ and $\widetilde{b}_{\gamma,\pm} = \widetilde{b}_{\gamma x} \pm i \widetilde{b}_{\gamma y}$, the spin blocks are given by
\begin{equation}
   F^{2\times 2}_{Z;ba} = \frac{1}{2} \begin{pmatrix} \widetilde{B}_z S_{ba} + \mu_{ba;\gamma} \widetilde{b}_{\gamma z} & \widetilde{B}_{-} S_{ba} + \mu_{ba;\gamma} \widetilde{b}_{\gamma,-} \\ \widetilde{B}_{+} S_{ba} + \mu_{ba;\gamma} \widetilde{b}_{\gamma,+} & \widetilde{B}_z S_{ba} + \mu_{ba;\gamma} \widetilde{b}_{\gamma z} \end{pmatrix}
\end{equation}

Turning to the exchange energy, we use Mulliken notation for the electron--electron repulsion integrals, and write
\begin{equation} \label{GHFexch}
\begin{split}
E_X & = -\frac{1}{2} \sum_{KL}^{\text{occ}} (KL|LK) = -\frac{1}{2} D^{\sigma\tau;da} D^{\tau\sigma;bc} (ab|cd),
\end{split}
\end{equation}
again with implicit summation over spin and basis function indices. The exchange contribution to the Fock matrix is obtained as
\begin{equation}
K_{fe}^{\tau\sigma} = \frac{\partial E_X}{\partial D^{\sigma\tau;ef}} = -D^{\tau\sigma;bc} (eb|cf).
\end{equation}
Alternatively, in $2\times 2$ component form,
\begin{equation}
K^{2\times 2}_{fe} = -\begin{pmatrix} D^{\uparrow\uparrow;bc} & D^{\uparrow\downarrow;bc} \\ D^{\downarrow\uparrow;bc} & D^{\downarrow\downarrow;bc} \end{pmatrix} (eb|cf).
\end{equation}

\section{Direction of second-order spin effects on $\mathcal{A}$} \label{SpinEffectsOnA}
\label{spindirection}

Because the space of RHF wave functions is a strict subset of the space of GHF wave functions, the corresponding ground state energies must satisfy $E_{\text{RHF}}(\mathbf{B},\mathbf{C}) \geq E_{\text{GHF}}(\mathbf{B},\mathbf{C})$. For any system without a permanent anapole moment and with a singlet ground state in the absence of magnetic fields, the second-order expansion in Eq.~\eqref{eqEnergyExpansion} yields
\begin{equation}
   E_{\text{RHF}}(\mathbf{0},\mathbf{C}) = E_0 - \frac{1}{2} \mathbf{C}^T \mathcal{A}_{\text{RHF}} \mathbf{C} \geq E_{\text{GHF}}(\mathbf{0},\mathbf{C}) = E_0 - \frac{1}{2} \mathbf{C}^T \mathcal{A}_{\text{GHF}} \mathbf{C},
\end{equation}
where $E_0 = E_{\text{RHF}}(\mathbf{0},\mathbf{0}) = E_{\text{GHF}}(\mathbf{0},\mathbf{0})$. The above inequality holds for all directions of $\mathbf{C}$ and it therefore follows that the difference
\begin{equation}
   \mathcal{A}_{\text{GHF}} - \mathcal{A}_{\text{RHF}} \geq 0
\end{equation}
is positive semidefinite. Moreover, when the corresponding eigenvalues are placed in ascending order, $\alpha_{\text{RHF},1} \leq \alpha_{\text{RHF},2} \leq \alpha_{\text{RHF},3}$ and $\alpha_{\text{GHF},1} \leq \alpha_{\text{GHF},2} \leq \alpha_{\text{GHF},3}$, it follows that $\alpha_{\text{GHF},j} \geq \alpha_{\text{RHF},j}$ for each $j=1,2,3$. All closed-shell molecules considered in Sec.~\ref{results} below exhibit a type of generalized orbital diamagnetism in the sense that $-\mathcal{A}_{\text{RHF}} > 0$ is positive definite. Hence, the RHF energy increases with increasing magnitude of $\mathbf{C}$. In these closed-shell molecules, the orbital and spin effects oppose each other, since the latter always exhibit a type of generalized paramagnetism in the sense that they lower the second-order energy.

\section{Additivity of second-order orbital and spin effects} 
\label{SpinandOrb}

For a fixed wave function, the orbital and spin contributions to the expectation value of the anapole moment are clearly additive. A more involved case is when a RHF ground state, optimized without any spin-Zeeman interactions, is allowed to relax into a GHF ground state subject to spin-Zeeman interaction. The orbital and spin contributions to the anapole susceptibility $\mathcal{A}$ remain additive in this case, despite the coupling of orbital and spin degrees of freedom. However, in the absence of a permanent GHF spin anapole moment, the second-order effect on the energy is exactly {\it half} the spin-Zeeman interaction. Hence,
\begin{equation}
   \label{eqQuadZeeman}
E_{\text{GHF}}(\mathbf{B},\mathbf{C}) - E_{\text{RHF}}(\mathbf{B},\mathbf{C}) = \frac{1}{2} E_{Z}^{\text{quadratic}} + \mathcal{O}(|\mathbf{C}|^3).
\end{equation}
One can also consider relaxation from an initial UHF state, optimized without spin-Zeeman interaction, to a GHF state. Taking into account the possibility of a permanent anapole moment, the corresponding relation is
\begin{equation}
   \label{eqLinQuadZeeman}
E_{\text{GHF}}(\mathbf{B},\mathbf{C}) - E_{\text{UHF}}(\mathbf{B},\mathbf{C}) =  E_{Z}^{\text{linear}} + \frac{1}{2} E_{Z}^{\text{quadratic}} + \mathcal{O}(|\mathbf{C}|^3).
\end{equation}
This effect is illustrated by our results in Sec.~\ref{results}. Below, we also provide a theoretical derivation.

The additivity indicated above can be understood as a general feature of perturbation theory, though it takes a surprising form in the present setting where non-perturbative numerical results are matched to a perturbation expansion with the spin-Zeeman term as the perturbation. A simple model case is provided by full diagonalization of a Hamiltonian in a space of of two $N$-electron states $\ket{0}$ and $\ket{1}$, chosen as eigenstates of a spin-independent Hamiltonian $\op{H}_0$. With a suitable choice of zero level for the energy, we can write the matrix representation as
\begin{equation}
   \mathbf{H}_0 = \begin{pmatrix} 0 & 0 \\ 0 & \omega \end{pmatrix}.
\end{equation}
For simplicity, assume $\mathbf{B} = 0$ and $\mathbf{C} = C \mathbf{e}_z$, and take the matrix elements related to the $z$-component of the {\it spin} anapole moment to be
\begin{equation}
   \mathbf{Z} = \begin{pmatrix} \bra{0} \op{a}_z \ket{0} & \bra{0} \op{a}_z \ket{1} \\ \bra{1} \op{a}_z \ket{0} & \bra{1} \op{a}_z \ket{1} \end{pmatrix} = \begin{pmatrix} \alpha_0 & \mu \\ \mu^* & \alpha_1 \end{pmatrix}.
\end{equation}
Diagonalization of $\mathbf{H}_0 - \tfrac{1}{2} C \mathbf{Z}$ now yields the (unnormalized) lowest eigenstate
\begin{equation}
   \Psi = 
   \begin{pmatrix}
       -2\omega - \Delta\alpha C -\sqrt{(\Delta\alpha^2 + 16 |\mu|^2) C^2 - 4\omega \Delta\alpha C + 4 \omega^2} \\
       4 \mu C
   \end{pmatrix},
\end{equation}
where $\Delta\alpha = \alpha_1 - \alpha_0$ is the difference in permanent anapole moments. Separating contributions of different orders in $C$, we obtain the spin-Zeeman energy
\begin{equation}
    E_Z = \frac{ \Psi^{\dagger} (-\frac{1}{2}) C \mathbf{Z} \Psi}{\Psi^{\dagger} \Psi} = -\frac{1}{2} C \alpha_0 - \frac{2|\mu|^2 C^2}{\omega} - \frac{3 \Delta\alpha |\mu|^2 C^3}{2 \omega^2} + \mathcal{O}(C^4)
\end{equation}
and the total energy
\begin{equation}
    E = \frac{ \Psi^{\dagger} (\mathbf{H}_0 -\frac{1}{2} C \mathbf{Z}) \Psi}{\Psi^{\dagger} \Psi} = -\frac{1}{2} C \alpha_0 - \frac{|\mu|^2 C^2}{\omega} - \frac{\Delta\alpha |\mu|^2 C^3}{2 \omega^2} + \mathcal{O}(C^4).
\end{equation}
Hence, half the second-order spin-Zeeman interaction, and two thirds of the third-order interaction, is cancelled by the response in other degrees of freedom.

Our current problem in the Hartree--Fock setting differs from the model Hamiltonian in that the Fock operator is density- or state dependent and the ground state needs to be found through a self-consistent field procedure.
In order to provide an explanation for Eqs.~\eqref{eqQuadZeeman} and \eqref{eqLinQuadZeeman} tailored to the present Hartree--Fock setting, we provide an analysis using McWeeny's formalism~\cite{McWeeny1962,MCWEENY_BOOK_2001}. Assume for simplicity an orthonormal basis of 2-component spinors and let $\mathbf{D}$ denote the one-particle reduced density matrix. All GHF states give rise to idempotent density matrices, $\mathbf{D}^2 = \mathbf{D}$. Then the Fock matrix can be written as a sum of a density-independent and a density-dependent part,
\begin{equation} \label{FockDef}
  \mathbf{F} = \mathbf{h} + \mathbf{G}(\mathbf{D}). 
\end{equation}
The Hartree--Fock energy is conveniently expressed in terms of the modified matrix $\mathbf{F}'$:
\begin{equation}
   E = \trace{\mathbf{F}' \mathbf{D}}, \quad \text{with} \quad \mathbf{F}' = \mathbf{h} + \frac{1}{2} \mathbf{G}(\mathbf{D}),
\end{equation}
and the density-matrix form of the Roothaan--Hall equation is
\begin{equation}
  \label{eqFDcom}
  \mathbf{F} \mathbf{D} - \mathbf{D} \mathbf{F} = \mathbf{0}.
\end{equation}

Expanding in orders of the perturbation strength $\lambda$, we write F as
\begin{equation}
   \label{pert_expand}
   \mathbf{F} = \mathbf{F}^{(0)} + \lambda \mathbf{F}^{(1)} + \lambda^2 \mathbf{F}^{(2)} + \ldots
\end{equation}
and analogously $\mathbf{h}$, $\mathbf{G}$ and $\mathbf{D}$ as well as the energy $E$.

Using the projector $\mathbf{P} = \mathbf{D}^{(0)}$ onto the occupied space of the unperturbed solution and its complement $\mathbf{Q} = \mathbf{I} - \mathbf{P}$, a projector on the virtual space, any matrix can be decomposed into its four projected blocks $\mathbf{M} = \mathbf{M}_{\text{oo}} + \mathbf{M}_{\text{ov}} + \mathbf{M}_{\text{vo}} + \mathbf{M}_{\text{vv}}$, with $\mathbf{M}_{\text{oo}} = \mathbf{P} \mathbf{M} \mathbf{P}$, $\mathbf{M}_{\text{ov}} = \mathbf{P} \mathbf{M} \mathbf{Q}$, and so on.
Occupied-virtual projections of the perturbative expansion of the idempotency condition $\mathbf{D}^2 = \mathbf{D}$ enable us to relate density matrices of various orders with each other while Eq.~\eqref{eqFDcom} provide us with the equations to determine the density matrices at a given order.
For the details of this procedure we refer to the paper by McWeeny~\cite{McWeeny1962}.
The final expressions for the first order density matrices are given by,
\begin{align}
\mathbf{D}_{\text{oo}}^{(1)} &= \mathbf{D}_{\text{vv}}^{(1)} = \mathbf{0}, \\
\mathbf{D}_{\text{ov}}^{(1)} &= \mathbf{f}_0^{-1} \mathbf{F}_{\text{ov}}^{(1)} + \mathbf{f}_0^{-1} \mathbf{D}_{\text{ov}}^{(1)} \mathbf{f}_0 \label{Dov_eq}
\end{align}
where $\mathbf{f}_0 = \mathbf{F}^{(0)}$ to simplify the notation. Iteration of Eq.~\eqref{Dov_eq} yields
\begin{eqnarray}
  \mathbf{D}_{\text{ov}}^{(1)} = \sum_{m=0}^{\infty} \mathbf{f}_0^{-(m+1)} \mathbf{F}_{\text{ov}}^{(1)} \mathbf{f}_0^m \label{CPHF}.
\end{eqnarray}
It should be noted that this is a form of the coupled perturbed HF self-consistency condition~\cite{MCWEENY_BOOK_2001}, since $\mathbf{F}^{(1)}_{\text{ov}}$ depends on $\mathbf{D}^{(1)}_{\text{ov}}$.
In our finite-field computation we do not explicitly solve this equation but an expansion of our final density in orders of perturbation would yield a first order contribution satisfying this equation. For our present purposes, it is sufficient to establish the above relation to eliminate $\mathbf{F}^{(1)}_{\text{ov}}$, and we do not need to further determine $\mathbf{D}^{(1)}_{\text{ov}}$.

Collecting the first order contributions to the energy, we find
\begin{equation}
   E^{(1)} 
            = \trace{ \mathbf{D}^{(0)} \mathbf{h}^{(1)}} + \trace{ \mathbf{D}^{(1)} \mathbf{F}^{(0)}}.
 \label{E1_1}
\end{equation}
The last term vanishes, since $\mathbf{F}^{(0)}$ can be chosen diagonal and $\mathbf{D}^{(1)}$ then only has off-diagonal components---hence, $\trace{\mathbf{D}^{(1)} \mathbf{F}^{(0)}} = 0$ and the equation simplifies to $E^{(1)} = \trace{\mathbf{D}^{(0)} \mathbf{h}^{(1)}}$.

Turning to the second-order contributions, we first note that the idempotency condition determines the blocks $\mathbf{D}^{(2)}_{\text{oo}}$ and $\mathbf{D}^{(2)}_{\text{vv}}$:
\begin{align}
   \label{den2ndorder}
  \mathbf{D}_{\text{oo}}^{(2)} & = -\mathbf{D}_{\text{ov}}^{(1)} \mathbf{D}_{\text{vo}}^{(1)},   \\
  \mathbf{D}_{\text{vv}}^{(2)} & = \mathbf{D}_{\text{vo}}^{(1)} \mathbf{D}_{\text{ov}}^{(1)}.
   \label{den2ndorderVV}
\end{align}
The remaining blocks $\mathbf{D}^{(2)}_{\text{ov}}$ and $\mathbf{D}^{(2)}_{\text{vo}}$ are not needed to determine the second-order contribution to the energy.
We may now write
\begin{equation}
   E^{(2)} = \trace{ \mathbf{F}'^{(2)} \mathbf{D}^{(0)} + \mathbf{F}'^{(1)} \mathbf{D}^{(1)} + \mathbf{F}'^{(0)} \mathbf{D}^{(2)} }.
\end{equation}
To evaluate $\mathbf{F}^{(0)} \mathbf{D}^{(2)}$, we note that since $\mathbf{F}^{(0)}$ is diagonal (in the basis of canonical orbitals), only the diagonal components of $\mathbf{D}^{(2)}$ contribute to the trace.
Following McWeeny, we use first Eqs.~\eqref{den2ndorder} and \eqref{den2ndorderVV} and then cyclic permutation on Eq.~\eqref{CPHF} to arrive at the relation
\begin{equation}
   \trace{ \mathbf{f}_0 (\mathbf{D}_{\text{oo}}^{(2)} + \mathbf{D}_{\text{vv}}^{(2)}) } = - \trace{ \sum_{n=0}^{\infty} \mathbf{F}_{\text{ov}}^{(1)} \mathbf{f}_0^{n} \mathbf{F}_{\text{vo}}^{(1)} \mathbf{f}_0^{-(n+1)}} = - \trace{ \mathbf{F}_{\text{ov}}^{(1)} \mathbf{D}_{\text{vo}}^{(1)} } = - \trace{ \mathbf{F}_{\text{vo}}^{(1)} \mathbf{D}_{\text{ov}}^{(1)} },
\end{equation}
where we again use the shorthand notation $\mathbf{f}_0 = \mathbf{F}^{(0)}$. Thus,
\begin{equation}
   \trace{ \mathbf{F}^{(0)} \mathbf{D}^{(2)} } = - \trace{ \mathbf{F}_{\text{ov}}^{(1)} \mathbf{D}_{\text{vo}}^{(1)} } = - \trace{ \mathbf{F}_{\text{vo}}^{(1)} \mathbf{D}_{\text{ov}}^{(1)} } = - \frac{1}{2} \trace{ \mathbf{F}^{(1)} \mathbf{D}^{(1)} }.
\end{equation}
and, consequently,
\begin{equation}
   \label{eqEOrder2}
   E^{(2)} = \trace{ \mathbf{h}^{(2)} \mathbf{D}^{(0)} + \tfrac{1}{2} \mathbf{h}^{(1)} \mathbf{D}^{(1)} }
\end{equation}

Next, we choose $\lambda$ to be a formal scaling factor applied to the spin-Zeeman interaction. More specifically, the $\lambda=0$ reference is a UHF ground state (RHF $\subset$ UHF) obtained without a spin-Zeeman term. The $\lambda=1$ case, with full spin-Zeeman interaction, corresponds to a GHF solution. Explicitly,
\begin{equation}
   \mathbf{h} = \mathbf{h}^{(0)}  + \lambda \mathbf{h}^{(1)}, \quad \text{with} \quad \op{h}^{(1)} = \sum_l \mathbf{B}_{\text{tot}}(\mathbf{r}_l)\cdot\op{\mathbf{S}}_l.
\end{equation}
There are no higher order corrections to $\mathbf{h}$, e.g., $\mathbf{h}^{(2)} = \mathbf{0}$. However, the perturbation of $\mathbf{h}^{(0)}$ perturbs the self-consistency condition and therefore induces corrections to the energy, density matrix and Fock matrix to all orders in $\lambda$. Using Eqs.~\eqref{E1_1} and \eqref{eqEOrder2} above, the difference between the GHF energy $E_{\text{GHF}} = E$ and the UHF energy $E_{\text{UHF}} = E^{(0)}$ can now be written
\begin{equation}
   E_{\text{GHF}} - E_{\text{UHF}} = \lambda \trace{\mathbf{D}^{(0)} \mathbf{h}^{(1)}} + \frac{\lambda^2}{2} \trace{\mathbf{D}^{(1)} \mathbf{h}^{(1)}} + \mathcal{O}(\lambda^3 \|\mathbf{h}^{(1)}\|^3),
\end{equation}
where $\|\mathbf{h}^{(1)}\|$ is the largest eigenvalue of $\mathbf{h}^{(1)}$.
Setting $\lambda = 1$ finally shows that only half the spin-Zeeman interaction for the induced spin density is present in the total $E_{\text{GHF}}$. The other half is cancelled by the coupling between the spin and orbital degrees of freedom.

\section{Results and Discussion} \label{results}

The implementation described in Sec.~\ref{GHFtheory} was applied to study a set of small molecules subject to magnetic field gradients.
The H$_2$ molecule was used for a preliminary study of the effect of transverse field gradients on closed shell molecules. The O$_2$ molecule was used as the corresponding exemple of an open-shell system. 
Further studies on the induced anapole moments were carried out on H$_2$O$_2$ and CHFXY (X,Y=Cl,Br).
The GHF model was used to capture joint spin and orbital effects, and the RHF model was used to isolate the purely orbital effects in closed shell systems. For the open shell molecule studied, the UHF model, with the spin-Zeeman term disabled, was used for the same purpose.

All calculations were performed using the {\sc London} program~\cite{Tellgren2008,LondonProgram}, to which we have added the necessary GHF implementation. The anapole susceptibilities $\mathcal{A}$ and $\mathcal{M}$ were calculated by second-order numerical differentiation of the energies. The susceptibilities $\mathcal{A}'$, $\mathcal{M}'$, and $\mathcal{M}''$ were calculated by a mixed approach: expectation values approximating first order analytic derivatives were differentiated once numerically. The symmetric finite difference formula was used, with step sizes of $\epsilon =0.01$a.u. for $\mathbf{B}$ and $\epsilon'=0.005$~a.u. for $\mathbf{C}$.
The smaller value for $\epsilon'$ was chosen as the effect of $\mathbf{C}$ on the local magnetic field is scaled by the interatomic distances in the molecule.
The error in the energy is quadratic in the step size within the limits to which the energy is converged while the error in the analytically computed moments (first derivative of energy) is linear.
All numerical results presented in this paper are given in SI-based atomic units---see earlier work for the conversion factors to SI units~\cite{Tellgren2013}.

The basis sets employed for studying basis set convergence come from the family of Dunning's correlation consistent basis sets~\cite{DUNNING_JCP90_1989,WOON_JCP100_2975} with and without augmentation with diffuse functions. 
The names of the basis sets are prefixed with `L' to denote the use of London atomic orbitals and `u' to indicate that the basis sets are uncontracted. 
The studies of the energies of H$_2$ and O$_2$ in a range of non-uniform fields, are carried out using the Luaug-cc-pVQZ basis set.
The location of the gauge origin, $\mathbf{g}$, only affects the non-LAO calculations. 
For H$_2$ and O$_2$, $\mathbf{g}$ is on one of the H or O atoms while for H$_2$O$_2$ it is at the mid-point of the O-O bond.
For the CHFXY group of molecules, $\mathbf{g}$ is on the central C atom.
Different reference points $\mathbf{h}$ for the linear component of the field have been explored.
The equilibrium bond lengths (R$_{\text{eq}}$) for H$_2$ and O$_2$ were taken to be 1.3984 a.u. and 2.287 a.u., respectively. Geometries for the other molecules are as reported in an earlier publication~\cite{Tellgren2013} and are also provided in the Supplementary Information.

\subsection{H$_2$}

\begin{figure}[]
\centering
\includegraphics[width=0.9\linewidth]{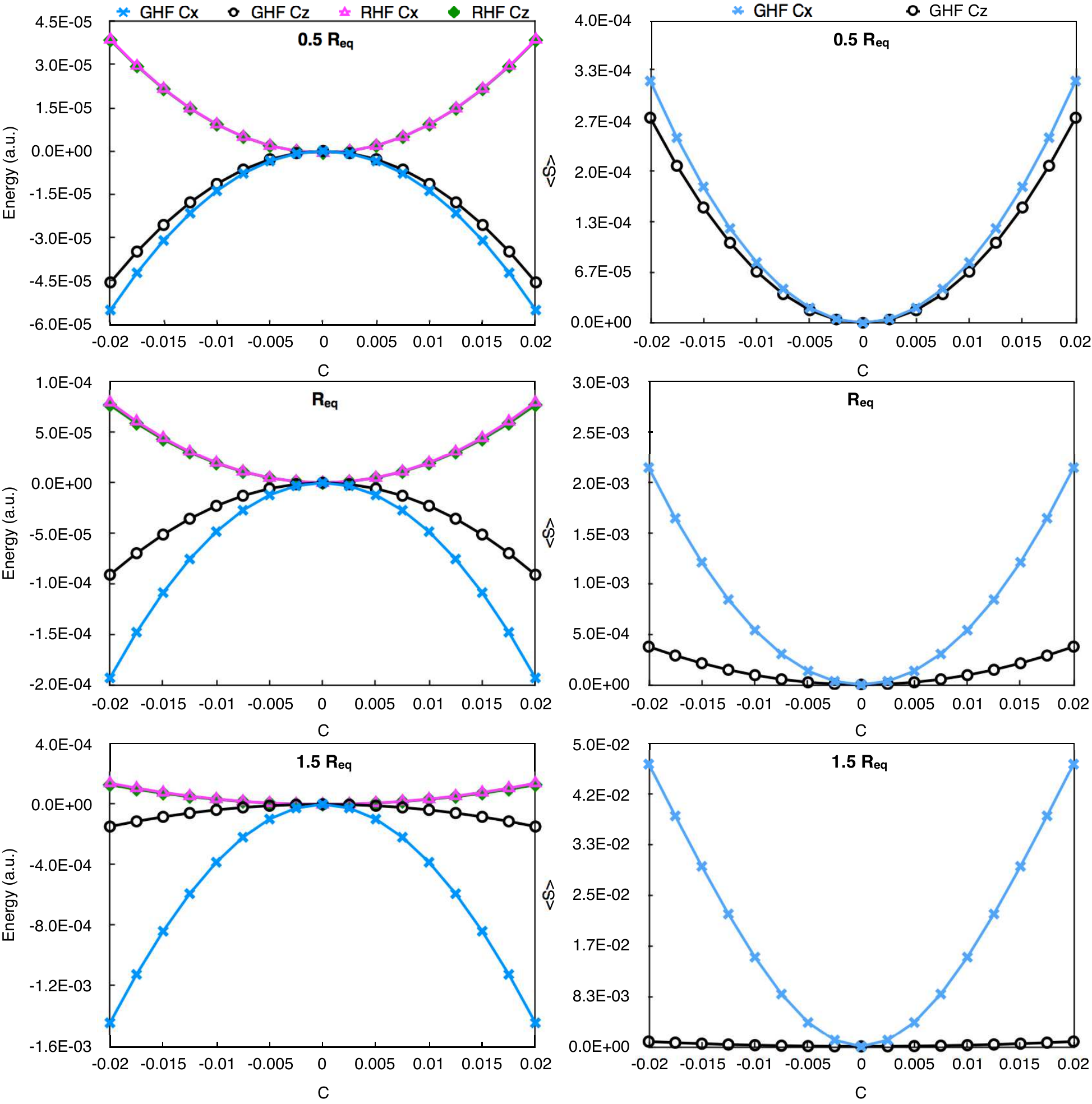}
\caption[CvarH2]{H$_2$: The first column of plots shows the change in energy of H$_2$ (Luaug-cc-pVQZ) with variation in the gradient of the external non-uniform magnetic field, $\mathbf{C}=C_x\mathbf{e}_x+C_y\mathbf{e}_y+C_z\mathbf{e}_z$.The second column of plots show the corresponding changes in the total spin quantum number, $S$.}
\label{fig:cvariationh2}
\end{figure}

For the H$_2$ molecule, the leading order response of the energy is quadratic in $\mathbf{C}$, as expected from any system that is a closed-shell singlet in the absence of magnetic fields.
In RHF computations where the response is restricted to orbital effects, the energy increases sharply in a parabolic curve.
In the case of GHF, the spin degrees of freedom allow a negation of the orbital effects leading to partial or full cancellation or even an inversion of the parabolic energy change.
For H$_2$, the balance of spin and orbital effects is seen to depend neatly on the bond length (see Fig. \ref{fig:cvariationh2}).
In our calculations, the H-H bond axis is aligned to the $z$-axis and
we observe identical spin and orbital effects for $C_x$ and $C_y$.
Thus, only the variation with $C_x$ is plotted.
The reference point, $\mathbf{h}$, is placed at the centre of the H-H bond at each of the bond lengths we have studied.
At $R = 0.5 R_{\text{eq}}$, the GHF plot is inverted relative to the RHF indicating a dominance of the spin effect introduced by mixing of the triplet state with the singlet ground state of H$_2$ and this trend continues as we stretch to $R = 1.5 R_{\text{eq}}$, where the downward curvature of the GHF plot is even larger.

The total spin quantum number, which directly indicates the amount of spin-breaking induced by the external magnetic field, is plotted as a function of $C$ at various points on the potential energy surface of H$_2$ in the second column of Fig.~\ref{fig:cvariationh2}. This quantity is obtained from the relation
\begin{equation}
    S(S+1) = \langle \op{S}^2 \rangle,
\end{equation}
and despite the terminology it is no longer a good quantum number.
The spin-breaking is found to be directly proportional to $C$ at all geometries.

\begin{figure}
\centering
\includegraphics[width=0.9\linewidth]{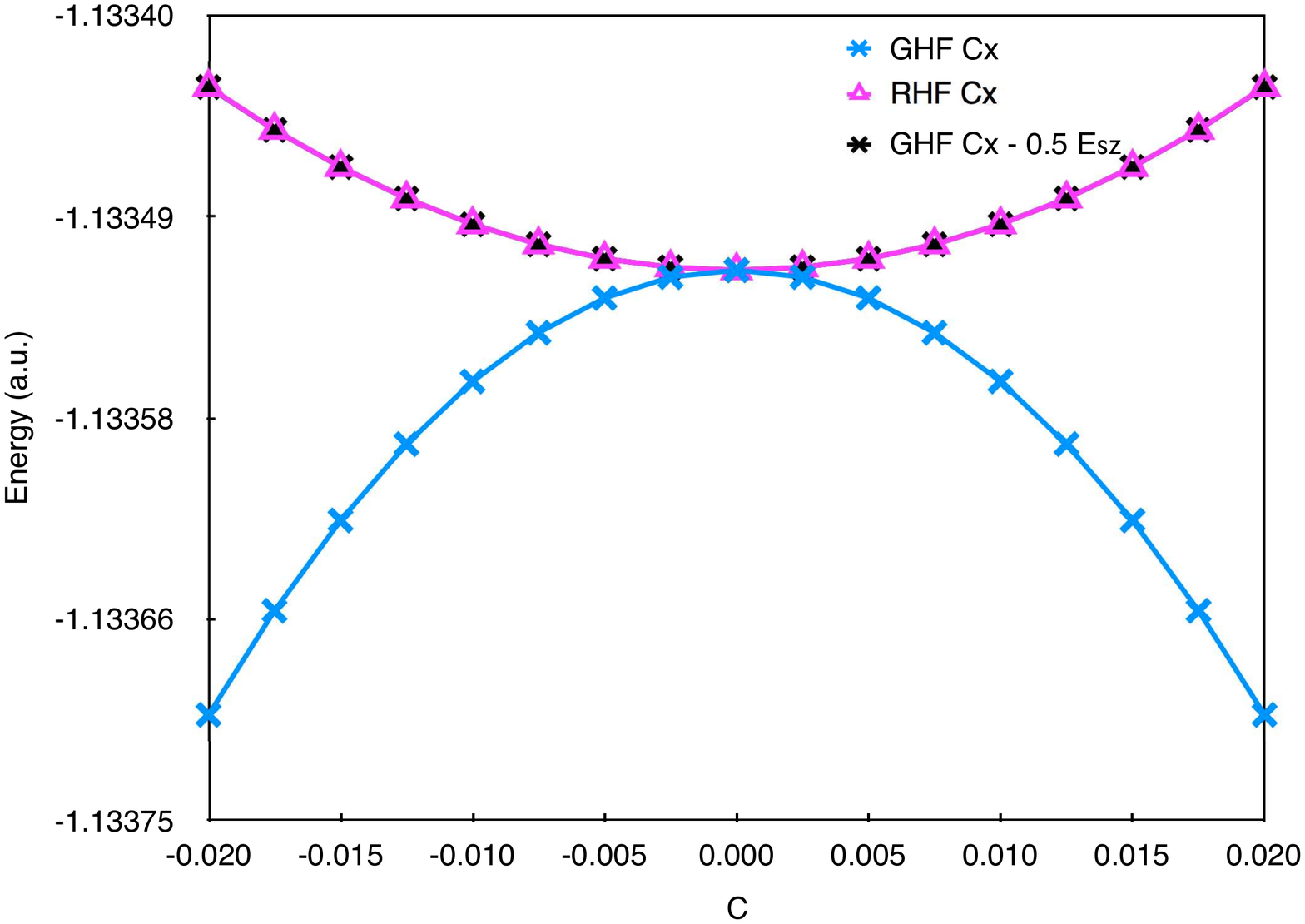}
\caption{H$_2$: The plot numerically demonstrates that the spin-Zeeman and orbital-Zeeman contributions to the total energy are not additive but are such that the lowering of the energy by the spin-Zeeman term is offset by exactly half of its value by the orbital term. This holds generally to second order.}
\label{fig:cvariationh2additivity}
\end{figure}

\begin{figure}[]
	\centering
	\includegraphics[width=0.9\linewidth]{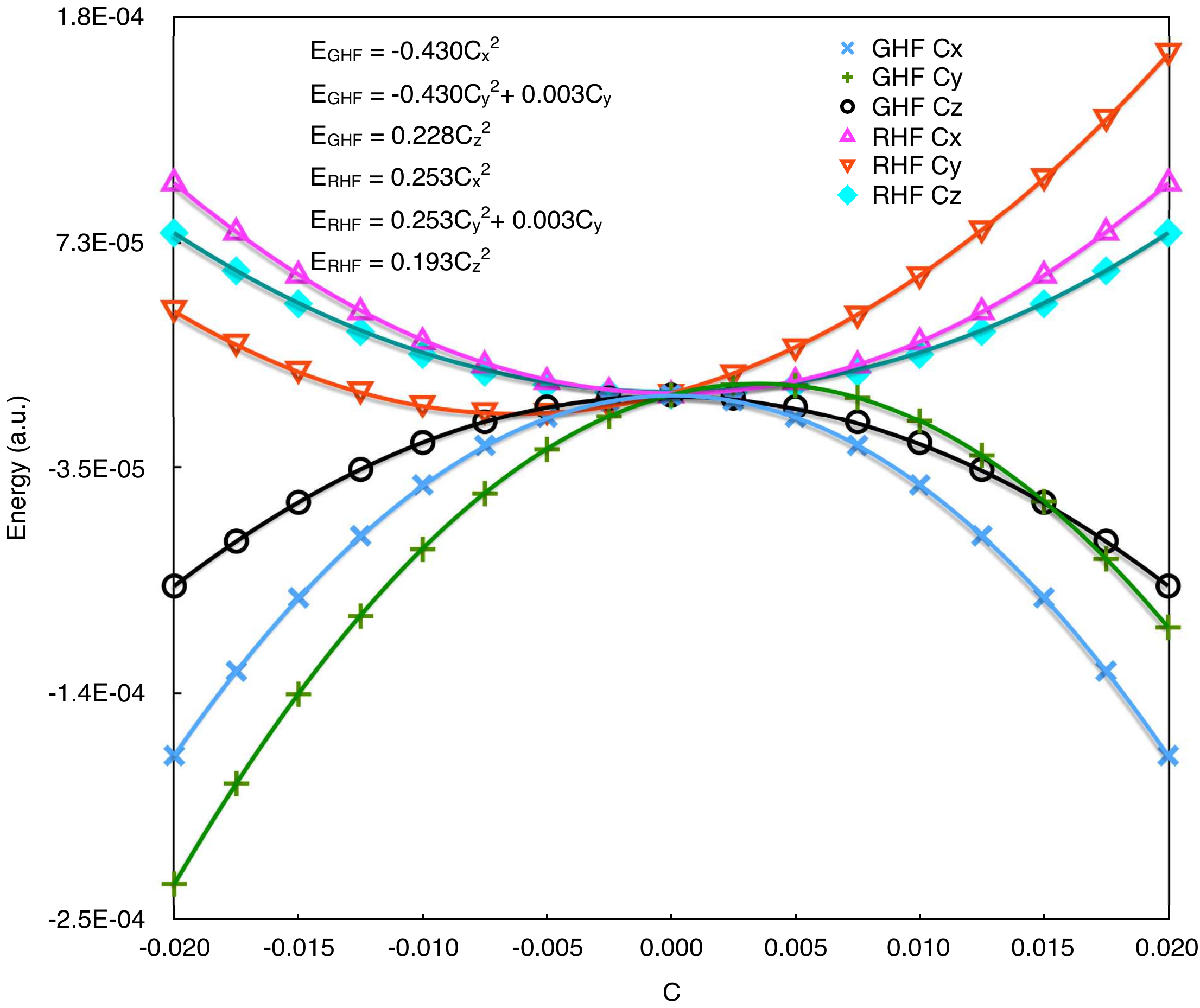}
	\caption[CvarWithBH2]{H$_2$: The plot shows the change in energy of H$_2$ (Luaug-cc-pVQZ) with variation in the gradient of the external non-uniform magnetic field, $\mathbf{C}=C_x\mathbf{e}_x+C_y\mathbf{e}_y+C_z\mathbf{e}_z$, in the presence of a constant uniform field, $\mathbf{B}=0.01\textbf{e}_x$. The reference point for the gradient of the field, $\mathbf{h}$, is placed on a H atom. The continuous lines are the polynomial fits to the data points, whose equations are indicated on the plot.}
	\label{fig:cvariationBh2}
\end{figure}

Fig. \ref{fig:cvariationh2additivity} demonstrates the relation of the GHF and RHF energies and numerically verifies our assertion in Sec. \ref{SpinandOrb}. On shifting up the GHF energies by half of the spin-Zeeman energy, the plot comes to lie exactly on top of the RHF plot.

When both a uniform component, $\mathbf{B}$, and a transverse gradient, $\mathbf{C}$ with \textbf{h} on one of the H atoms, are switched on, the mixed anapole susceptibility is expected to come into play.
In Fig.~\ref{fig:cvariationBh2}, we plot the variation in energy of H$_2$ with the components C$_x$, C$_y$ and C$_z$, with a fixed uniform component B$_x$=0.01~a.u.
For the component C$_x$, the transverse gradient and uniform component are in the same direction and the mixed susceptibility tensors have zero diagonal elements resulting in a parabolic energy curve just like in the previous case of vanishing $\mathbf{B}$.
However, the $\mathbf{B} = 0$ and $\mathbf{B} = 0.01 \mathbf{e}_x$~a.u.\ curves do not coincide as the values of the local field, $\mathbf{B}_{\text{tot}}(\mathbf{r})$, are different.
The nature of the variation of energy with C$_y$, on the other hand, is modified in the presence of the B$_x$ component.
An induced anapole moment, a$_y$ = B$_x \mathcal{M}_{xy}$, leads to the addition of a linear component to the energy as a function of C$_y$. 
When RHF calculations are performed, the orbital effects result in an unsymmetric parabola.
On the other hand, the presence of spin-field interactions in the GHF case results in a change in the curvature accompanied by a flipping of the sign of the anapole susceptibility.
The linear component remains the same as the RHF calculation.
This is borne out by our analysis in Sec.~\ref{SpinandOrb} as the zeroth order density is the same for RHF and GHF.
When \textbf{h} is placed at the centre of the bond, the linear effects are cancelled on account of symmetry.

\subsection{O$_2$}

The O$_2$ molecule in its ground triplet state serves as our sample molecule for interaction of non-singlet molecules with non-uniform fields.
In order to isolate pure orbital effects, we carried out UHF calculations without the spin-Zeeman term for the $m_s=1$ triplet state. The advantage of this constraint on the UHF wave function, compared to the $m_s = 0$ constraint, is that it guarantees that the wave function optimization does not accidentally lead to a singlet state.
In our calculations, the O-O bond axis is aligned to the $z$-axis. 

\begin{figure}[]
	\centering
	\includegraphics[width=0.8\linewidth]{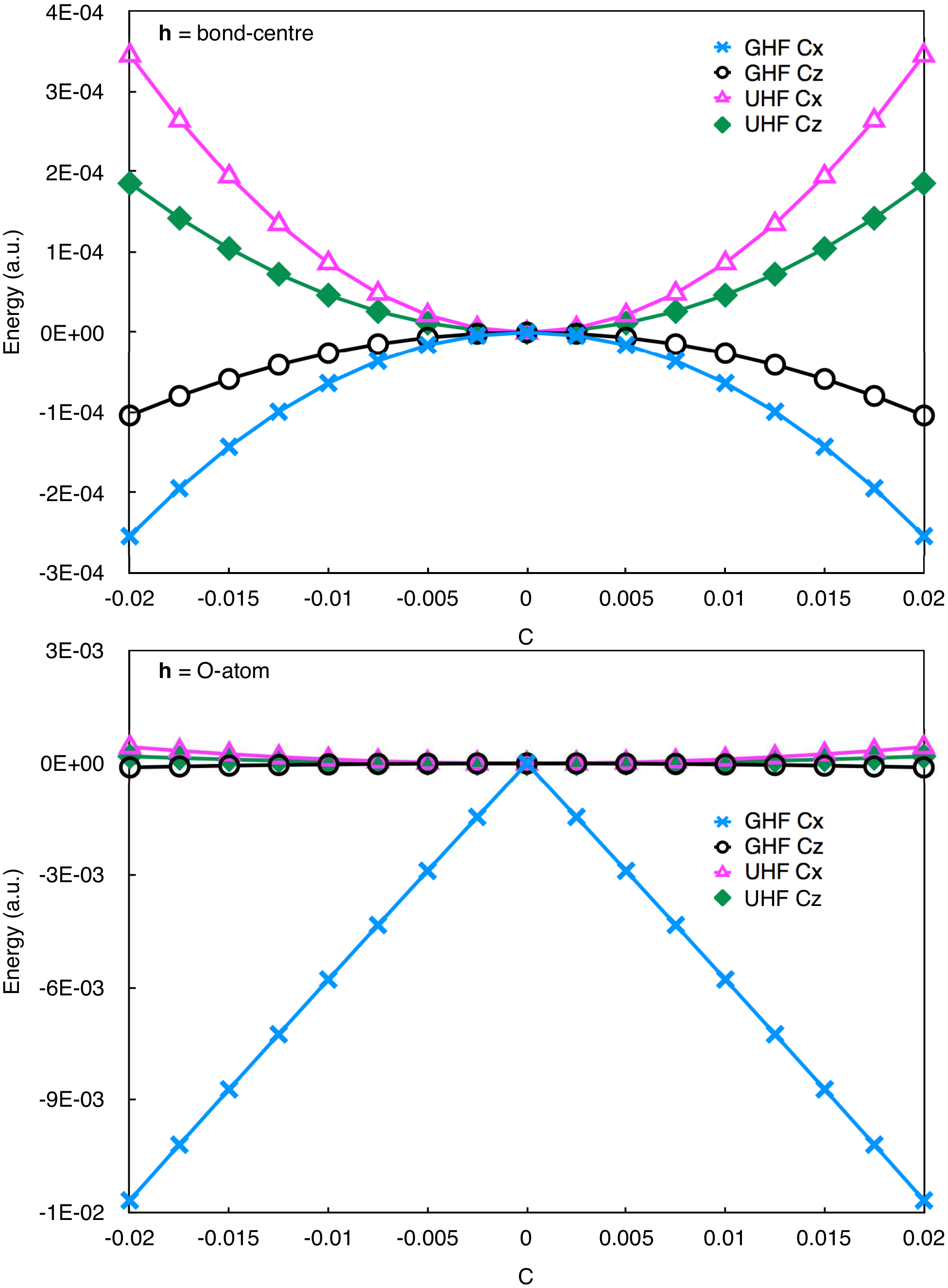}
	\caption[CvarO2]{O$_2$: The plot shows the change in energy of O$_2$ (Luaug-cc-pVQZ) as a function of transverse field gradient, $\mathbf{C}=C_x\mathbf{e}_x+C_y\mathbf{e}_y+C_z\mathbf{e}_z$. The top plot corresponds to the reference point for the gradient, $\mathbf{h}$, placed at the bond-centre and the bottom plot to $\mathbf{h}$ placed on one O atom. In the former case, first order effects are cancelled due to symmetry reasons.}
	\label{fig:cvariationO2}
\end{figure}

As shown in Fig.~\ref{fig:cvariationO2}, the energy vs.\ C$_z$ curve is flatter than the curves for the C$_x$ and C$_y$ components.
When the spin-Zeeman interactions are included in a GHF calculation, we find that the energy vs.\ C$_z$ curve changes only slightly.
The energy as a function of C$_x$ (or C$_y$), on the other hand, changes drastically.
In the top plot, \textbf{h} is placed at the centre of the O-O bond. Inversion symmetry causes the first order spin-Zeeman interaction with C$_x$ (or C$_y$) to cancel.
In the bottom plot, on the other hand, \textbf{h} is placed unsymmetrically on an O atom and first order effects are observed.
Due to the ground state degeneracy at $\mathbf{B} = \mathbf{C} = \mathbf{0}$, ground states with positive, negative, and vanishing permanent anapole moments a$_x$ and a$_y$ are possible. These states have different energy curves that cross at C$_x$ = 0, resulting in a cusp when the lowest energy is plotted as a function of C$_x$.
Hence, the first order spin-Zeeman interaction results in a sharp decrease in the energy as soon as the field is switched on.
A superposition of this linear variation with a parabolic response on account of the orbital effects in the opposite direction gives the bottom curve in Fig. \ref{fig:cvariationO2} its characteristic shape.

\begin{figure}[]
\centering
\includegraphics[width=0.7\linewidth]{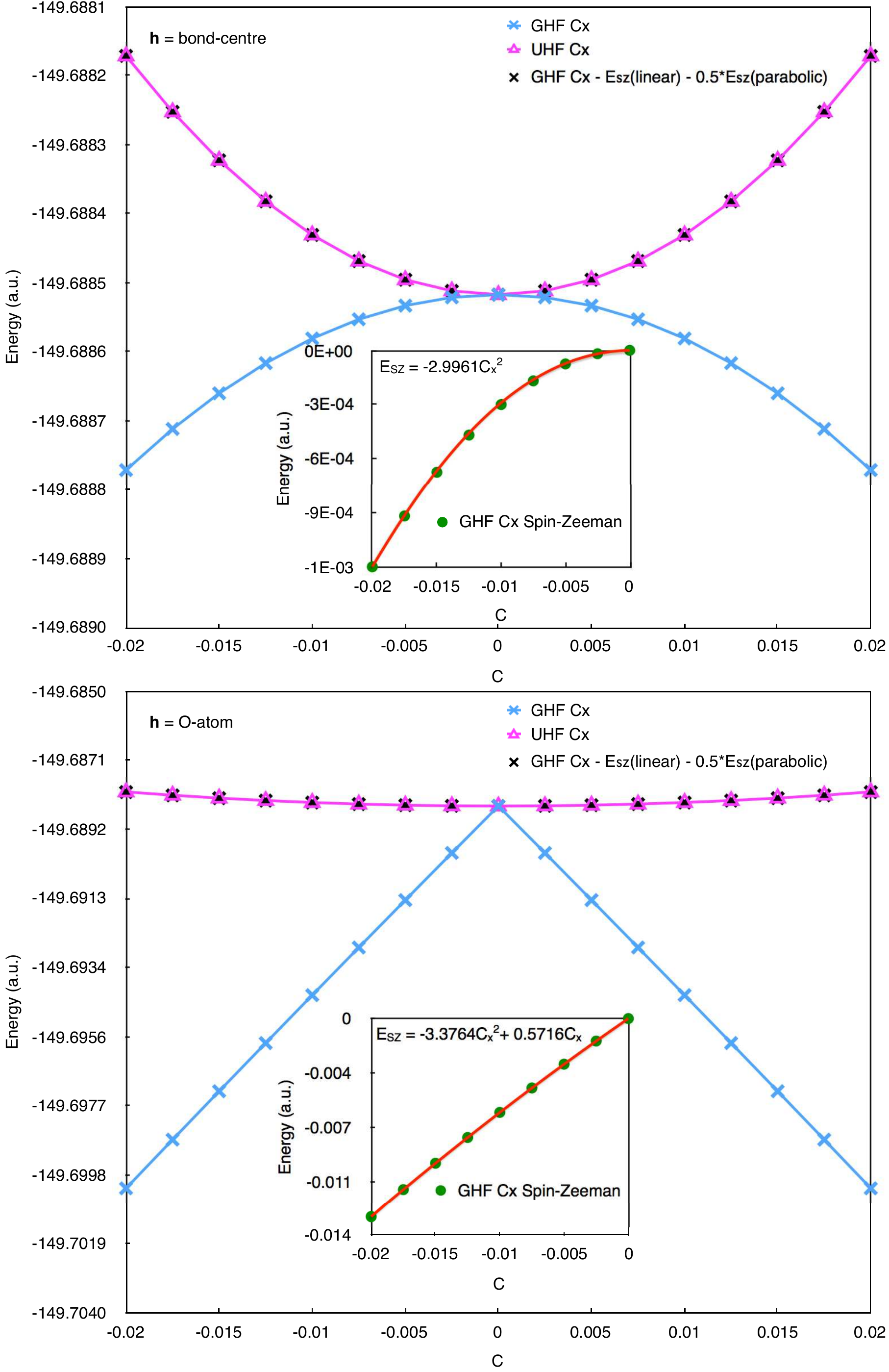}
\caption{O$_2$: The plots numerically demonstrate that the spin-Zeeman and orbital-Zeeman contributions to the total energy are not additive but are such that the lowering of the energy by the spin-Zeeman term is offset by exactly half of its value by the orbital term. The inset shows the variation of only the spin-Zeeman energy with $C$ and the fitting equation is indicated on the plot. The top plot corresponds to the reference point for the gradient, $\mathbf{h}$, placed at the bond-centre and the bottom plot to $\mathbf{h}$ placed on an O atom. In the former case, first order effects are cancelled due to symmetry reasons.}
\label{fig:cvariationo2additivity}
\end{figure}

Fig. ~\ref{fig:cvariationo2additivity} demonstrates the perturbative relation between UHF and GHF, as discussed in Sec. \ref{SpinandOrb} for open-shell molecules. The top plot corresponds to \textbf{h} placed at the centre of the O-O bond while in the bottom plot, \textbf{h} is placed on the O atom at the origin of the coordinate system.
The insets in Fig.~\ref{fig:cvariationo2additivity} are plots of the spin-Zeeman energy vs.\ C$_x$, which are fitted to separate out the linear and parabolic components of the spin-Zeeman interaction.
The fitting equation is also indicated in the graph.
In the top plot, the first order linear spin effects are cancelled due to the symmetry of the system but the main curve in the bottom plot clearly demonstrates the relation in Eq.~\eqref{eqLinQuadZeeman}. This indicates that our numerical computations are adequately described by up to first- and second-order effects on the energy. We note that in this study we have used relatively weaker fields up to 0.020 a.u. where higher order effects do not set in for small molecules like H$_2$ and O$_2$.

The basis set convergence of the anapole susceptibility values for triplet O$_2$ follow the same trends as the singlet molecules reported below.
However, the spin effects push the values in the same direction as the orbital effects unlike in the other cases---see Tables~\ref{Tab:O2UHF} and \ref{Tab:O2GHF}.
This is to be expected as O$_2$, unlike the singlet molecules, exhibits first order spin effects, i.e., permanent spin magnetic dipole and spin anapole moments, which dominate the energy response in the absence of symmetry reasons which may cancel these effects.
By contrast, the leading orbital effects are always second order in the field.

\begin{table}
	\renewcommand{\arraystretch}{0.7}
	\begin{tabular}{|c|c|c|c|c|c|c|}
		\hline 
		Basis			& \multicolumn{3}{c|}{$\mathcal{A}$} & \multicolumn{3}{c|}{$\mathcal{A'}$} \\
\hline
uSTO-3G &  -15.451 & N/A & N/A &  -15.438 &    0.000 &   -0.000 \\
& N/A &  -15.451 & N/A &   -0.000 &  -15.438 &   -0.000 \\
& N/A & N/A &   -2.692 &   -0.000 &   -0.000 &   -2.692 \\
\hline
ucc-pVDZ &  -10.094 & N/A & N/A &  -10.086 &   -0.000 &   -0.000 \\
& N/A &  -10.094 & N/A &    0.000 &  -10.086 &   -0.000 \\
& N/A & N/A &   -2.988 &   -0.000 &   -0.000 &   -2.988 \\
\hline
ucc-pVTZ &   -6.932 & N/A & N/A &   -6.925 &   -0.000 &   -0.000 \\
& N/A &   -6.932 & N/A &   -0.000 &   -6.925 &   -0.000 \\
& N/A & N/A &   -2.386 &   -0.000 &   -0.000 &   -2.386 \\
\hline
uaug-cc-pVDZ &   -5.710 & N/A & N/A &   -5.705 &   -0.000 &    0.000 \\
& N/A &   -5.710 & N/A &   -0.000 &   -5.705 &    0.000 \\
& N/A & N/A &   -2.244 &   -0.000 &   -0.000 &   -2.244 \\
\hline
uaug-cc-pVTZ &   -4.603 & N/A & N/A &   -4.597 &    0.000 &   -0.000 \\
& N/A &   -4.603 & N/A &   -0.000 &   -4.597 &   -0.000 \\
& N/A & N/A &   -1.941 &   -0.000 &   -0.000 &   -1.940 \\
\hline
uaug-cc-pVQZ &   -4.368 & N/A & N/A &   -4.363 &   -0.000 &   -0.000 \\
& N/A &   -4.368 & N/A &   -0.000 &   -4.363 &    0.000 \\
& N/A & N/A &   -1.859 &    0.000 &   -0.000 &   -1.859 \\
\hline
LuSTO-3G &   -4.423 & N/A & N/A &   -6.201 &    0.000 &   -0.000 \\
& N/A &   -4.423 & N/A &   -0.000 &   -6.201 &   -0.000 \\
& N/A & N/A &   -2.692 &    0.000 &    0.000 &   -2.692 \\
\hline
Lucc-pVDZ &   -5.225 & N/A & N/A &   -6.352 &    0.000 &   -0.000 \\
& N/A &   -5.225 & N/A &    0.000 &   -6.352 &   -0.000 \\
& N/A & N/A &   -2.988 &   -0.000 &    0.000 &   -2.988 \\
\hline
Lucc-pVTZ &   -4.950 & N/A & N/A &   -5.484 &   -0.000 &   -0.000 \\
& N/A &   -4.950 & N/A &    0.000 &   -5.484 &   -0.000 \\
& N/A & N/A &   -2.386 &    0.000 &    0.000 &   -2.386 \\ 
\hline       
Luaug-cc-pVDZ &   -4.697 & N/A & N/A &   -4.840 &    0.000 &    0.000 \\
& N/A &   -4.697 & N/A &   -0.000 &   -4.840 &    0.000 \\
& N/A & N/A &   -2.244 &   -0.000 &   -0.000 &   -2.244 \\
\hline   
Luaug-cc-pVTZ &   -4.427 & N/A & N/A &   -4.438 &    0.000 &   -0.000 \\
& N/A &   -4.427 & N/A &    0.000 &   -4.438 &   -0.000 \\
& N/A & N/A &   -1.941 &    0.000 &    0.000 &   -1.940 \\
\hline    
Luaug-cc-pVQZ &   -4.330 & N/A & N/A &   -4.328 &    0.000 &   -0.000 \\
& N/A &   -4.330 & N/A &   -0.000 &   -4.328 &    0.000 \\
& N/A & N/A &   -1.859 &    0.000 &   -0.000 &   -1.859 \\
\hline
	\end{tabular}
\caption{O$_2$ : Basis set convergence of the cartesian anapole susceptibility tensor computed at the UHF level. L = London atomic orbitals, u = uncontracted. $\mathcal{A}$ and $\mathcal{A'}$ are defined in Eqs.~\eqref{Asus_def} and \eqref{Aprimesus_def} respectively.}
 \label{Tab:O2UHF}
\end{table}

\begin{table}
	\renewcommand{\arraystretch}{0.7}
	\begin{tabular}{|c|c|c|c|c|c|c|}
		\hline 
		Basis			& \multicolumn{3}{c|}{$\mathcal{A}$} & \multicolumn{3}{c|}{$\mathcal{A'}$} \\
		\hline
		uSTO-3G &  -10.187 & N/A & N/A &  -10.121 &    0.000 &   -0.000 \\
		& N/A &  -10.187 & N/A &    0.000 &  -10.121 &   -0.000 \\
		& N/A & N/A &   -2.171 &   -0.000 &   -0.000 &   -2.171 \\
		\hline
		ucc-pVDZ &   -4.536 & N/A & N/A &   -4.475 &    0.000 &   -0.000 \\
		& N/A &   -4.536 & N/A &   -0.000 &   -4.475 &   -0.000 \\
		& N/A & N/A &   -1.840 &   -0.000 &   -0.000 &   -1.840 \\
		\hline
		ucc-pVTZ &   -0.807 & N/A & N/A &   -0.756 &    0.000 &    0.000 \\
		& N/A &   -0.807 & N/A &    0.000 &   -0.756 &    0.000 \\
		& N/A & N/A &   -0.470 &    0.000 &   -0.000 &   -0.471 \\
		\hline
		uaug-cc-pVDZ &    0.899 & N/A & N/A &    0.935 &   -0.000 &   -0.000 \\
		& N/A &    0.899 & N/A &   -0.000 &    0.935 &   -0.000 \\
		& N/A & N/A &    0.464 &   -0.000 &   -0.000 &    0.464 \\
		\hline
		uaug-cc-pVTZ &    2.129 & N/A & N/A &    2.160 &    0.000 &   -0.000 \\
		& N/A &    2.129 & N/A &   -0.000 &    2.160 &   -0.000 \\
		& N/A & N/A &    0.902 &   -0.000 &    0.000 &    0.902 \\
		\hline
		uaug-cc-pVQZ &    2.392 & N/A & N/A &    2.422 &    0.000 &    0.000 \\
		& N/A &    2.392 & N/A &    0.000 &    2.422 &   -0.000 \\
		& N/A & N/A &    1.035 &    0.000 &    0.000 &    1.035 \\
		\hline
		LuSTO-3G &    0.774 & N/A & N/A &   -0.968 &    0.000 &   -0.000 \\
		& N/A &    0.774 & N/A &   -0.000 &   -0.968 &   -0.000 \\
		& N/A & N/A &   -2.171 &   -0.000 &   -0.000 &   -2.171 \\
		\hline
		Lucc-pVDZ &    0.273 & N/A & N/A &   -0.819 &   -0.000 &   -0.000 \\
		& N/A &    0.273 & N/A &    0.000 &   -0.819 &   -0.000 \\
		& N/A & N/A &   -1.840 &   -0.000 &    0.000 &   -1.840 \\
		\hline
		Lucc-pVTZ &    1.136 & N/A & N/A &    0.636 &    0.000 &    0.000 \\
		& N/A &    1.136 & N/A &   -0.000 &    0.636 &    0.000 \\
		& N/A & N/A &   -0.470 &    0.000 &   -0.000 &   -0.471 \\
		\hline
		Luaug-cc-pVDZ &    1.897 & N/A & N/A &    1.780 &    0.000 &   -0.000 \\
		& N/A &    1.897 & N/A &    0.000 &    1.780 &   -0.000 \\
		& N/A & N/A &    0.464 &    0.000 &    0.000 &    0.464 \\
		\hline
		Luaug-cc-pVTZ &    2.302 & N/A & N/A &    2.316 &    0.000 &   -0.000 \\
		& N/A &    2.302 & N/A &    0.000 &    2.316 &   -0.000 \\
		& N/A & N/A &    0.902 &   -0.000 &    0.000 &    0.902 \\
		\hline
		Luaug-cc-pVQZ &    2.429 & N/A & N/A &    2.457 &    0.000 &    0.000 \\
		& N/A &    2.429 & N/A &   -0.000 &    2.457 &   -0.000 \\
		& N/A & N/A &    1.035 &    0.000 &    0.000 &    1.035 \\
		\hline
	\end{tabular} 
\caption{O$_2$ : Basis set convergence of the cartesian anapole susceptibility tensor computed at the GHF level. L = London atomic orbitals, u = uncontracted. $\mathcal{A}$ and $\mathcal{A'}$ are defined in Eqs.~\eqref{Asus_def} and \eqref{Aprimesus_def} respectively.}	
\label{Tab:O2GHF}
\end{table}

\subsection{H$_2$O$_2$}

\begin{figure}
\centering
\includegraphics[width=0.98\linewidth]{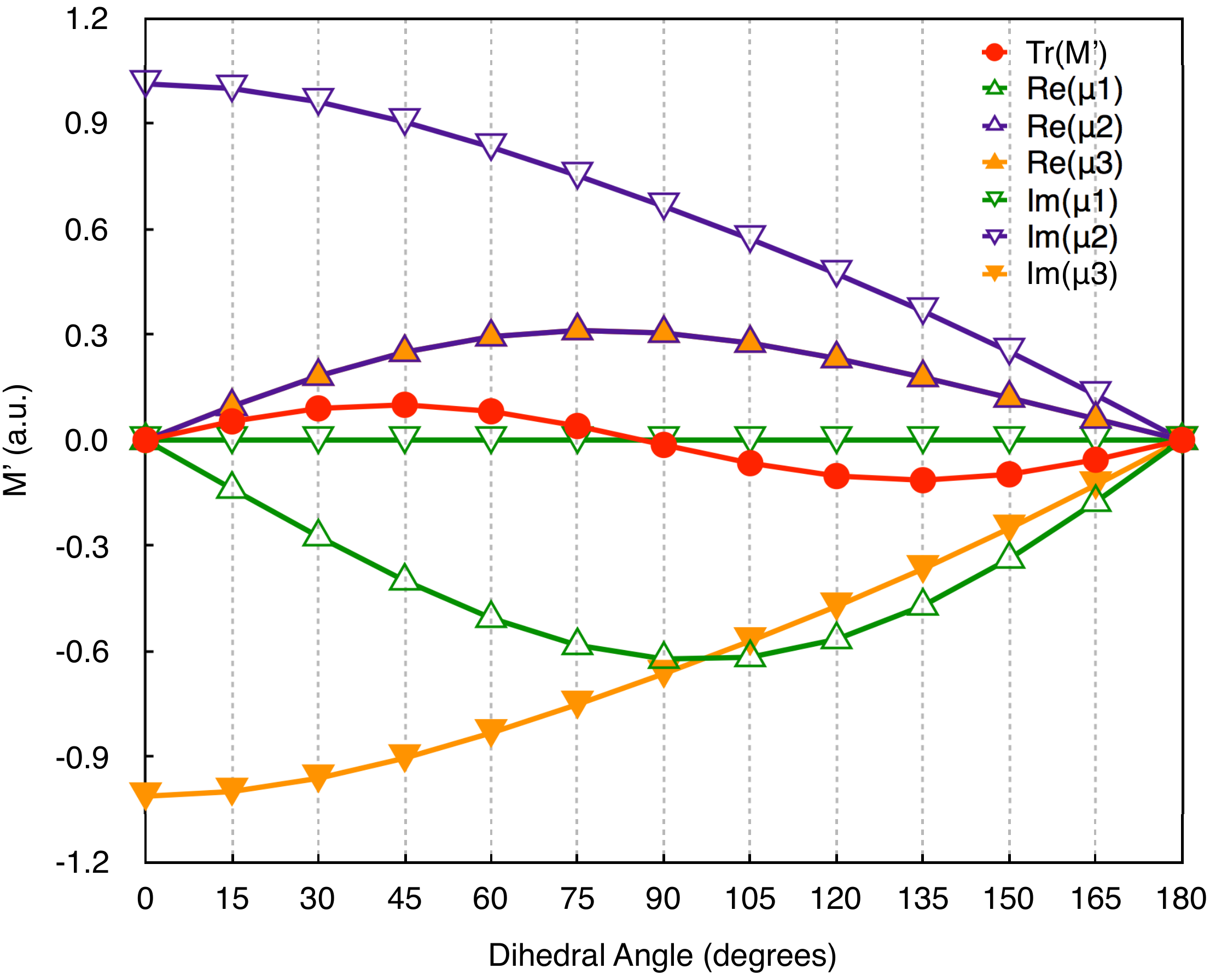}
\caption{H$_2$O$_2$: The plot shows the variation of the mixed anapole susceptibility, $\mathcal{M}$' with the dihedral angle. $\trace(\mathcal{M}')$ is the trace of the $\mathcal{M}$' tensor and {$\mu_i$}, $i=1,2,3$, are its eigenvalues. The trace goes through a sign shift at a dihedral angle of 86$^\circ$. Two of the eigenvalues $\mu_2$ and $\mu_3$ constitute complex-conjugated pairs while $\mu_1$ is real.}
\label{fig:anapoleplotsh2o2dihedral}
\end{figure}

A previous study of anapole moments~\cite{Tellgren2013} (see also a recent further analysis~\cite{PELLONI_JPCA121_9369}) focused on H$_2$O$_2$ as this system can be continuously deformed from an achiral to a chiral and back to an achiral structure by changing the dihedral angle continuously from 0$^\circ$ to 180$^\circ$, the energy minimum being at 120$^\circ$. This is useful since features of the mixed anapole susceptibility tensor, $\mathcal{M}$, are sensitive to chirality~\cite{Lazzeretti2014}. However, spin-symmetry breaking effects only enter as a correction to $\mathcal{M}$ that is second order in C$_x$. Hence, the GHF results in
Fig.~\ref{fig:anapoleplotsh2o2dihedral} are identical, to within numerical noise, to what was obtained in the RHF case~\cite{Tellgren2013}.
At a dihedral angle of 86$^{\circ}$ the trace undergoes a sign change.
This marks the point of highest chirality for H$_2$O$_2$.

At a fixed dihedral angle of 120$^\circ$, the full 3$\times$3 tensors for $\mathcal{A}$, $\mathcal{A}'$, $\mathcal{M}$, $\mathcal{M}'$ and $\mathcal{M}''$ have been computed using both RHF and GHF---see Tables~\ref{H2O2_RHF_A_con}-\ref{H2O2_GHF_A_uncon}. An earlier study~\cite{Tellgren2013} showed that LAOs dramatically accelerate basis set convergence of RHF level anapole susceptibilities. This remains true when spin effects are included.
However, LAOs also violate the assumptions of the Hellmann--Feynman theorem.
This manifests itself as a discrepancy between the values of $\mathcal{A}$ and $\mathcal{A}'$ in Tables~\ref{H2O2_RHF_A_con}-\ref{H2O2_GHF_A_uncon}.
A loss of the symmetry between $\mathcal{A}'_{yz}$ and $\mathcal{A}'_{zy}$ is also notable in the results. Both of these discrepancies vanish in the basis set limit. Conversely, they become larger for smaller basis sets.
Uncontracted basis sets improve the basis set convergence, which is natural given that the contraction coefficients were not optimized for the response to magnetic field gradients. A comparison of the results for contracted and uncontracted basis sets are reported in Table \ref{H2O2_RHF_A_con}-\ref{H2O2_RHF_A_uncon} (RHF level) as well as in Tables~\ref{H2O2_GHF_A_con}-\ref{H2O2_GHF_A_uncon} (GHF level).
Similar conclusions hold for the basis effects on the mixed anapole susceptibility tensors in Tables~\ref{H2O2_GHF_M_con} and \ref{H2O2_GHF_M_uncon}.
Coming to one of the main conclusions of this work, the spin contributions to the anapole susceptibility, $\mathcal{A}$, are found to be much larger than the orbital contributions and act in opposition to the orbital effects. This is in line with our theoretical understanding presented in Sec. \ref{SpinEffectsOnA}.
There are no spin effects on the mixed anapole susceptibilities and thus GHF and RHF results are identical and are not reported separately.
The mixed anapole susceptibilities computed using GHF are reported in Tables \ref{H2O2_GHF_M_con} and \ref{H2O2_GHF_M_uncon}. 

\begin{table}
	\renewcommand{\arraystretch}{0.7}
	\begin{tabular}{|c|c|c|c|c|c|c|}
		\hline 
		Basis			& \multicolumn{3}{c|}{$\mathcal{A}$} & \multicolumn{3}{c|}{$\mathcal{A'}$} \\
		\hline 
		STO-3G		&  -14.076 &   -0.000 &   -0.000 &  -14.076 &   -0.000 &   -0.000 \\
		&   -0.000 &  -12.778 &    1.904 &   -0.000 &  -12.778 &    1.905 \\
		&   -0.000 &    1.904 &   -6.955 &   -0.000 &    1.905 &   -6.955 \\
		\hline 
		cc-pVDZ 	&   -7.448 &   -0.000 &   -0.000 &   -7.448 &   -0.000 &    0.000 \\
		&   -0.000 &   -7.792 &    0.995 &    0.000 &   -7.793 &    0.995 \\
		&   -0.000 &    0.995 &   -4.458 &   -0.000 &    0.995 &   -4.458 \\
		\hline 
		cc-pVTZ 	&   -4.164 &   -0.000 &   -0.000 &   -4.165 &    0.000 &    0.000 \\
		&   -0.000 &   -5.213 &    0.700 &    0.000 &   -5.213 &    0.700 \\
		&   -0.000 &    0.700 &   -3.381 &   -0.000 &    0.700 &   -3.381 \\
		\hline 
		aug-cc-pVDZ &   -3.657 &   -0.000 &   -0.000 &   -3.657 &   -0.000 &   -0.000 \\
		&   -0.000 &   -5.090 &    0.517 &    0.000 &   -5.090 &    0.517 \\
		&   -0.000 &    0.517 &   -2.869 &    0.000 &    0.517 &   -2.869 \\
		\hline 
		aug-cc-pVTZ &   -2.457 &   -0.000 &   -0.000 &   -2.458 &   -0.000 &    0.000 \\
		&   -0.000 &   -4.102 &    0.504 &   -0.000 &   -4.102 &    0.504 \\
		&   -0.000 &    0.504 &   -2.451 &   -0.000 &    0.504 &   -2.451 \\
		\hline 
		aug-cc-pVQZ &   -2.126 &   -0.000 &   -0.000 &   -2.127 &   -0.000 &   -0.000 \\
		&   -0.000 &   -3.803 &    0.489 &    0.000 &   -3.803 &    0.489 \\
		&   -0.000 &    0.489 &   -2.355 &    0.000 &    0.489 &   -2.356 \\
		\hline 
		LSTO-3G 	&   -3.597 &   -0.000 &   -0.000 &   -6.044 &   -0.000 &   -0.000 \\
		&   -0.000 &   -4.785 &    0.631 &   -0.000 &   -6.578 &    1.087 \\
		&   -0.000 &    0.631 &   -4.403 &    0.000 &    0.745 &   -5.207 \\
		\hline 
		Lcc-pVDZ 	&   -3.075 &   -0.000 &   -0.000 &   -4.244 &   -0.000 &   -0.000 \\
		&   -0.000 &   -4.514 &    0.718 &    0.000 &   -5.329 &    0.894 \\
		&   -0.000 &    0.718 &   -3.884 &   -0.000 &    0.711 &   -4.091 \\
		\hline 
		Lcc-pVTZ 	&   -2.627 &   -0.000 &   -0.000 &   -3.058 &    0.000 &   -0.000 \\
		&   -0.000 &   -4.093 &    0.615 &   -0.000 &   -4.372 &    0.683 \\
		&   -0.000 &    0.615 &   -3.207 &    0.000 &    0.619 &   -3.283 \\
		\hline 
		Laug-cc-pVDZ &   -2.375 &   -0.000 &   -0.000 &   -2.503 &   -0.000 &   -0.000 \\
		&   -0.000 &   -4.008 &    0.493 &    0.000 &   -4.087 &    0.504 \\
		&   -0.000 &    0.493 &   -2.810 &    0.000 &    0.469 &   -2.813 \\
		\hline 
		Laug-cc-pVTZ &   -2.118 &   -0.000 &   -0.000 &   -2.139 &    0.000 &   -0.000 \\
		&   -0.000 &   -3.797 &    0.491 &    0.000 &   -3.813 &    0.496 \\
		&   -0.000 &    0.491 &   -2.434 &    0.000 &    0.494 &   -2.440 \\
		\hline 
		Laug-cc-pVQZ &   -2.051 &   -0.000 &   -0.000 &   -2.053 &    0.000 &    0.000 \\
		&   -0.000 &   -3.750 &    0.485 &    0.000 &   -3.752 &    0.487 \\
		&   -0.000 &    0.485 &   -2.349 &    0.000 &    0.486 &   -2.352 \\					
		\hline 
	\end{tabular} 
	\caption{H$_2$O$_2$ at a dihedral angle of 120$^\circ$: Basis set convergence of the cartesian anapole susceptibility tensor computed with RHF. L = London atomic orbitals. Contracted basis sets have been used. $\mathcal{A}$ and $\mathcal{A'}$ are defined in Eqs.~\eqref{Asus_def} and \eqref{Aprimesus_def}, respectively.}
	\label{H2O2_RHF_A_con}
\end{table}  

\begin{table} 
	\renewcommand{\arraystretch}{0.7}
	\begin{tabular}{|c|c|c|c|c|c|c|}
		\hline 
		Basis			& \multicolumn{3}{c|}{$\mathcal{A}$} & \multicolumn{3}{c|}{$\mathcal{A'}$} \\
		\hline
		uSTO-3G &  -11.459 &   -0.000 &   -0.000 &  -11.459 &   -0.000 &    0.000 \\
		&   -0.000 &  -10.588 &    1.793 &   -0.000 &  -10.588 &    1.793 \\
		&   -0.000 &    1.793 &   -5.407 &    0.000 &    1.793 &   -5.407 \\
		\hline
		ucc-pVDZ &   -6.328 &   -0.000 &   -0.000 &   -6.328 &   -0.000 &    0.000 \\
		&   -0.000 &   -6.754 &    0.986 &   -0.000 &   -6.754 &    0.986 \\
		&   -0.000 &    0.986 &   -4.390 &    0.000 &    0.986 &   -4.389 \\
		\hline
		ucc-pVTZ &   -3.715 &   -0.000 &   -0.000 &   -3.715 &   -0.000 &   -0.000 \\
		&   -0.000 &   -4.810 &    0.700 &   -0.000 &   -4.810 &    0.700 \\
		&   -0.000 &    0.700 &   -3.347 &    0.000 &    0.700 &   -3.347 \\
		\hline
		uaug-cc-pVDZ &   -2.907 &   -0.000 &   -0.000 &   -2.908 &    0.000 &   -0.000 \\
		&   -0.000 &   -4.353 &    0.533 &    0.000 &   -4.353 &    0.533 \\
		&   -0.000 &    0.533 &   -2.869 &    0.000 &    0.533 &   -2.870 \\
		\hline
		uaug-cc-pVTZ &   -2.204 &   -0.000 &   -0.000 &   -2.204 &   -0.000 &    0.000 \\
		&   -0.000 &   -3.876 &    0.509 &    0.000 &   -3.877 &    0.509 \\
		&   -0.000 &    0.509 &   -2.464 &    0.000 &    0.509 &   -2.464 \\
		\hline
		uaug-cc-pVQZ &   -2.067 &   -0.000 &   -0.000 &   -2.066 &   -0.000 &    0.000 \\
		&   -0.000 &   -3.765 &    0.490 &   -0.000 &   -3.764 &    0.490 \\
		&   -0.000 &    0.490 &   -2.358 &   -0.000 &    0.490 &   -2.357 \\
		\hline
		LuSTO-3G &   -2.835 &   -0.000 &   -0.000 &   -5.231 &   -0.000 &   -0.000 \\
		&   -0.000 &   -4.257 &    0.825 &    0.000 &   -5.917 &    1.357 \\
		&   -0.000 &    0.825 &   -3.586 &   -0.000 &    0.840 &   -4.200 \\
		\hline
		Lucc-pVDZ &   -3.025 &   -0.000 &   -0.000 &   -4.107 &    0.000 &    0.000 \\
		&   -0.000 &   -4.436 &    0.718 &   -0.000 &   -5.177 &    0.905 \\
		&   -0.000 &    0.718 &   -3.839 &    0.000 &    0.718 &   -4.052 \\
		\hline
		Lucc-pVTZ &   -2.596 &   -0.000 &   -0.000 &   -3.011 &    0.000 &   -0.000 \\
		&   -0.000 &   -4.067 &    0.611 &   -0.000 &   -4.336 &    0.683 \\
		&   -0.000 &    0.611 &   -3.170 &    0.000 &    0.613 &   -3.246 \\
		\hline
		Luaug-cc-pVDZ &   -2.400 &   -0.000 &   -0.000 &   -2.507 &   -0.000 &   -0.000 \\
		&   -0.000 &   -4.014 &    0.501 &   -0.000 &   -4.070 &    0.514 \\
		&   -0.000 &    0.501 &   -2.814 &   -0.000 &    0.484 &   -2.820 \\
		\hline
		Luaug-cc-pVTZ &   -2.124 &   -0.000 &   -0.000 &   -2.140 &   -0.000 &    0.000 \\
		&   -0.000 &   -3.809 &    0.492 &   -0.000 &   -3.820 &    0.498 \\
		&   -0.000 &    0.492 &   -2.441 &   -0.000 &    0.495 &   -2.449 \\
		\hline
		Luaug-cc-pVQZ &   -2.053 &   -0.000 &   -0.000 &   -2.054 &   -0.000 &   -0.000 \\
		&   -0.000 &   -3.752 &    0.485 &   -0.000 &   -3.752 &    0.487 \\
		&   -0.000 &    0.485 &   -2.352 &   -0.000 &    0.486 &   -2.353 \\
		\hline 
	\end{tabular} 
	\caption{H$_2$O$_2$ at a dihedral angle of 120$^\circ$: Basis set convergence of the cartesian anapole susceptibility tensor computed with RHF. L = London atomic orbitals. Uncontracted basis sets have been used. $\mathcal{A}$ and $\mathcal{A'}$ are defined in Eqs.~\eqref{Asus_def} and \eqref{Aprimesus_def}, respectively.}
	\label{H2O2_RHF_A_uncon}
\end{table} 

\begin{table}
	\renewcommand{\arraystretch}{0.7}
	\begin{tabular}{|c|c|c|c|c|c|c|}
		\hline 
		Basis			& \multicolumn{3}{c|}{$\mathcal{A}$} & \multicolumn{3}{c|}{$\mathcal{A'}$} \\
		\hline
		STO-3G &   25.222 &   -0.000 &   -0.000 &   22.285 &    0.000 &    0.000 \\
		&   -0.000 &   25.304 &    2.338 &   -0.000 &   22.349 &    2.507 \\
		&   -0.000 &    2.338 &   -4.496 &    0.000 &    2.338 &   -4.496 \\
		\hline
		cc-pVDZ &   36.145 &   -0.000 &   -0.000 &   33.106 &    0.000 &    0.000 \\
		&   -0.000 &   34.841 &    0.249 &   -0.000 &   31.787 &    0.166 \\
		&   -0.000 &    0.249 &   -1.108 &    0.000 &    0.249 &   -1.108 \\
		\hline
		cc-pVTZ &   40.102 &   -0.000 &   -0.000 &   37.048 &    0.000 &    0.000 \\
		&   -0.000 &   38.160 &   -0.202 &   -0.000 &   35.093 &   -0.317 \\
		&   -0.000 &   -0.202 &    1.205 &   -0.000 &   -0.202 &    1.205 \\
		\hline
		aug-cc-pVDZ &   38.472 &   -0.000 &   -0.000 &   36.083 &   -0.000 &    0.000 \\
		&   -0.000 &   36.258 &   -0.229 &    0.000 &   33.857 &   -0.302 \\
		&   -0.000 &   -0.229 &    2.487 &    0.000 &   -0.229 &    2.487 \\
		\hline
		aug-cc-pVTZ &   42.735 &   -0.000 &   -0.000 &   39.627 &    0.000 &   -0.000 \\
		&   -0.000 &   40.321 &   -0.314 &    0.000 &   37.201 &   -0.420 \\
		&   -0.000 &   -0.314 &    2.995 &    0.000 &   -0.314 &    2.995 \\
		\hline
		aug-cc-pVQZ &   44.005 &   -0.000 &   -0.000 &   40.657 &   -0.000 &    0.000 \\
		&   -0.000 &   41.564 &   -0.341 &   -0.000 &   38.204 &   -0.458 \\
		&   -0.000 &   -0.341 &    3.110 &   -0.000 &   -0.342 &    3.110 \\
		\hline
		LSTO-3G &   35.702 &   -0.000 &   -0.000 &   30.318 &    0.000 &   -0.000 \\
		&   -0.000 &   33.299 &    1.062 &    0.000 &   28.552 &    1.690 \\
		&   -0.000 &    1.062 &   -1.944 &   -0.000 &    1.178 &   -2.748 \\
		\hline
		Lcc-pVDZ &   40.517 &   -0.000 &   -0.000 &   36.309 &   -0.000 &    0.000 \\
		&   -0.000 &   38.119 &   -0.029 &    0.000 &   34.250 &    0.065 \\
		&   -0.000 &   -0.029 &   -0.533 &    0.000 &   -0.036 &   -0.741 \\
		\hline
		Lcc-pVTZ &   41.639 &   -0.000 &   -0.000 &   38.154 &   -0.000 &   -0.000 \\
		&   -0.000 &   39.281 &   -0.286 &    0.000 &   35.934 &   -0.334 \\
		&   -0.000 &   -0.286 &    1.379 &   -0.000 &   -0.283 &    1.303 \\
		\hline
		Laug-cc-pVDZ &   39.753 &   -0.000 &   -0.000 &   37.237 &   -0.000 &   -0.000 \\
		&   -0.000 &   37.340 &   -0.253 &   -0.000 &   34.861 &   -0.314 \\
		&   -0.000 &   -0.253 &    2.546 &   -0.000 &   -0.277 &    2.543 \\
		\hline
		Laug-cc-pVTZ &   43.074 &   -0.000 &   -0.000 &   39.946 &   -0.000 &    0.000 \\
		&   -0.000 &   40.626 &   -0.327 &   -0.000 &   37.490 &   -0.428 \\
		&   -0.000 &   -0.327 &    3.012 &   -0.000 &   -0.324 &    3.006 \\
		\hline
		Laug-cc-pVQZ &   44.080 &   -0.000 &   -0.000 &   40.731 &    0.000 &    0.000 \\
		&   -0.000 &   41.618 &   -0.346 &   -0.000 &   38.256 &   -0.461 \\
		&   -0.000 &   -0.346 &    3.116 &   -0.000 &   -0.345 &    3.114 \\
		\hline 
	\end{tabular}
	\caption{H$_2$O$_2$ at a dihedral angle of 120$^\circ$: Basis set convergence of the cartesian anapole susceptibility tensor computed with GHF. L = London atomic orbitals. Contracted basis sets have been used. $\mathcal{A}$ and $\mathcal{A'}$ are defined in Eqs.~\eqref{Asus_def} and \eqref{Aprimesus_def}, respectively.}
	\label{H2O2_GHF_A_con}
\end{table}

\begin{table} 
	\renewcommand{\arraystretch}{0.7}
	\begin{tabular}{|c|c|c|c|c|c|c|}
		\hline 
		Basis			& \multicolumn{3}{c|}{$\mathcal{A}$} & \multicolumn{3}{c|}{$\mathcal{A'}$} \\
		\hline
		uSTO-3G &   20.713 &   -0.000 &   -0.000 &   19.668 &    0.000 &    0.000 \\
		&   -0.000 &   20.279 &    1.902 &   -0.000 &   19.220 &    1.944 \\
		&   -0.000 &    1.902 &   -2.337 &    0.000 &    1.902 &   -2.337 \\
		\hline
		ucc-pVDZ &   34.771 &   -0.000 &   -0.000 &   32.312 &    0.000 &   -0.000 \\
		&   -0.000 &   33.366 &    0.213 &   -0.000 &   30.892 &    0.140 \\
		&   -0.000 &    0.213 &   -0.808 &    0.000 &    0.213 &   -0.808 \\
		\hline
		ucc-pVTZ &   40.237 &   -0.000 &   -0.000 &   37.261 &    0.000 &    0.000 \\
		&   -0.000 &   38.260 &   -0.203 &   -0.000 &   35.271 &   -0.316 \\
		&   -0.000 &   -0.203 &    1.263 &   -0.000 &   -0.203 &    1.263 \\
		\hline
		uaug-cc-pVDZ &   38.251 &   -0.000 &   -0.000 &   36.059 &   -0.000 &    0.000 \\
		&   -0.000 &   36.032 &   -0.195 &   -0.000 &   33.828 &   -0.260 \\
		&   -0.000 &   -0.195 &    2.489 &   -0.000 &   -0.195 &    2.488 \\
		\hline
		uaug-cc-pVTZ &   43.429 &   -0.000 &   -0.000 &   40.210 &    0.000 &   -0.000 \\
		&   -0.000 &   40.985 &   -0.304 &    0.000 &   37.753 &   -0.413 \\
		&   -0.000 &   -0.304 &    2.985 &    0.000 &   -0.305 &    2.985 \\
		\hline
		uaug-cc-pVQZ &   44.071 &   -0.000 &   -0.000 &   40.722 &    0.000 &    0.000 \\
		&   -0.000 &   41.609 &   -0.341 &   -0.000 &   38.248 &   -0.457 \\
		&   -0.000 &   -0.341 &    3.110 &   -0.000 &   -0.341 &    3.109 \\
		\hline
		LuSTO-3G &   29.337 &   -0.000 &   -0.000 &   25.897 &   -0.000 &   -0.000 \\
		&   -0.000 &   26.611 &    0.934 &   -0.000 &   23.892 &    1.507 \\
		&   -0.000 &    0.934 &   -0.516 &    0.000 &    0.949 &   -1.130 \\
		\hline
		Lucc-pVDZ &   38.072 &   -0.000 &   -0.000 &   34.531 &    0.000 &   -0.000 \\
		&   -0.000 &   35.684 &   -0.056 &   -0.000 &   32.469 &    0.060 \\
		&   -0.000 &   -0.056 &   -0.257 &   -0.000 &   -0.055 &   -0.471 \\
		\hline
		Lucc-pVTZ &   41.356 &   -0.000 &   -0.000 &   37.966 &   -0.000 &   -0.000 \\
		&   -0.000 &   39.003 &   -0.292 &    0.000 &   35.746 &   -0.333 \\
		&   -0.000 &   -0.292 &    1.440 &   -0.000 &   -0.289 &    1.364 \\
		\hline
		Luaug-cc-pVDZ &   38.758 &   -0.000 &   -0.000 &   36.460 &   -0.000 &   -0.000 \\
		&   -0.000 &   36.370 &   -0.228 &   -0.000 &   34.111 &   -0.279 \\
		&   -0.000 &   -0.228 &    2.544 &   -0.000 &   -0.245 &    2.538 \\
		\hline
		Luaug-cc-pVTZ &   43.509 &   -0.000 &   -0.000 &   40.274 &    0.000 &    0.000 \\
		&   -0.000 &   41.053 &   -0.322 &    0.000 &   37.810 &   -0.424 \\
		&   -0.000 &   -0.322 &    3.009 &    0.000 &   -0.318 &    3.001 \\
		\hline
		Luaug-cc-pVQZ &   44.085 &   -0.000 &   -0.000 &   40.734 &    0.000 &   -0.000 \\
		&   -0.000 &   41.623 &   -0.345 &    0.000 &   38.259 &   -0.460 \\
		&   -0.000 &   -0.345 &    3.116 &   -0.000 &   -0.344 &    3.113 \\
		\hline 
	\end{tabular}
	\caption{H$_2$O$_2$ at a dihedral angle of 120$^\circ$: Basis set convergence of the cartesian anapole susceptibility tensor computed with GHF. L = London atomic orbitals. Uncontracted basis sets have been used. $\mathcal{A}$ and $\mathcal{A'}$ are defined in Eqs.~\eqref{Asus_def} and \eqref{Aprimesus_def}, respectively.}
	\label{H2O2_GHF_A_uncon}
\end{table} 

\begin{table}
	\renewcommand{\arraystretch}{0.7}
	\begin{tabular}{|c|c|c|c|c|c|c|c|c|c|}
		\hline 
		Basis			& \multicolumn{3}{c|}{$\mathcal{M}$} &\multicolumn{3}{c|}{$\mathcal{M'}$} & \multicolumn{3}{c|}{$\mathcal{M''}$} \\
		\hline
		STO-3G &    0.600 &   -0.000 &   -0.000 &    0.634 &    0.000 &   -0.000 &    0.602 &    0.000 &    0.000 \\
		&   -0.000 &   -0.749 &   -1.643 &    0.000 &   -0.786 &   -1.643 &   -0.000 &   -0.752 &   -1.643 \\
		&   -0.000 &    2.052 &   -0.012 &   -0.000 &    2.040 &   -0.012 &   -0.000 &    2.049 &   -0.012 \\
		\hline
		cc-pVDZ &    0.141 &   -0.000 &   -0.000 &    0.172 &   -0.000 &   -0.000 &    0.142 &   -0.000 &    0.000 \\
		&   -0.000 &   -0.193 &   -0.702 &    0.000 &   -0.228 &   -0.702 &   -0.000 &   -0.195 &   -0.702 \\
		&   -0.000 &    0.926 &   -0.018 &    0.000 &    0.917 &   -0.019 &   -0.000 &    0.925 &   -0.019 \\
		\hline
		cc-pVTZ &   -0.306 &   -0.000 &   -0.000 &   -7.858 &   -0.000 &   -0.000 &   -0.305 &   -0.000 &    0.000 \\
		&   -0.000 &    0.249 &    1.528 &   -0.000 &   -8.804 &    1.531 &   -0.000 &    0.248 &   -0.540 \\
		&   -0.000 &    0.756 &   -0.350 &   -0.000 &   -8.552 &   -0.350 &   -0.000 &    0.755 &   -0.025 \\
		\hline
		aug-cc-pVDZ &   -0.403 &   -0.000 &   -0.000 &   -0.377 &    0.000 &    0.000 &   -0.403 &    0.000 &    0.000 \\
		&   -0.000 &    0.349 &   -0.450 &    0.000 &    0.319 &   -0.450 &    0.000 &    0.348 &   -0.450 \\
		&   -0.000 &    0.715 &   -0.033 &   -0.000 &    0.708 &   -0.033 &    0.000 &    0.714 &   -0.033 \\
		\hline
		aug-cc-pVTZ &   -0.520 &   -0.000 &   -0.000 &   -8.182 &    0.000 &    0.000 &   -0.519 &   -0.000 &   -0.000 \\
		&   -0.000 &    0.458 &    1.898 &    0.000 &   -8.356 &    1.900 &    0.000 &    0.456 &   -0.434 \\
		&   -0.000 &    0.704 &   -0.314 &    0.000 &   -8.809 &   -0.315 &    0.000 &    0.703 &   -0.034 \\
		\hline
		aug-cc-pVQZ &   -0.529 &   -0.000 &   -0.000 &   -8.158 &    0.000 &    0.000 &   -0.528 &    0.000 &    0.000 \\
		&   -0.000 &    0.462 &    1.874 &    0.000 &   -8.448 &    1.875 &   -0.000 &    0.460 &   -0.424 \\
		&   -0.000 &    0.698 &   -0.314 &    0.000 &   -8.861 &   -0.315 &   -0.000 &    0.697 &   -0.035 \\
		\hline
		LSTO-3G &   -0.620 &   -0.000 &   -0.000 &   -0.562 &   -0.000 &   -0.000 &   -0.459 &   -0.000 &   -0.000 \\
		&   -0.000 &    0.524 &   -0.501 &    0.000 &    0.468 &   -0.503 &    0.000 &    0.288 &   -0.930 \\
		&   -0.000 &    0.838 &   -0.033 &   -0.000 &    0.472 &   -0.009 &    0.000 &    1.375 &   -0.025 \\
		\hline
		Lcc-pVDZ &   -0.583 &   -0.000 &   -0.000 &   -0.555 &    0.000 &    0.000 &   -0.276 &    0.000 &    0.000 \\
		&   -0.000 &    0.539 &   -0.341 &   -0.000 &    0.507 &   -0.348 &    0.000 &    0.210 &   -0.569 \\
		&   -0.000 &    0.617 &   -0.062 &    0.000 &    0.542 &   -0.046 &    0.000 &    0.771 &   -0.026 \\
		\hline
		Lcc-pVTZ &   -0.557 &   -0.000 &   -0.000 &   -8.197 &   -0.000 &   -0.000 &   -0.440 &    0.000 &   -0.000 \\
		&   -0.000 &    0.505 &    1.617 &   -0.000 &   -8.433 &    1.641 &   -0.000 &    0.373 &   -0.511 \\
		&   -0.000 &    0.656 &   -0.373 &   -0.000 &   -8.702 &   -0.361 &   -0.000 &    0.724 &   -0.028 \\
		\hline
		Laug-cc-pVDZ &   -0.590 &   -0.000 &   -0.000 &   -0.556 &    0.000 &    0.000 &   -0.511 &    0.000 &   -0.000 \\
		&   -0.000 &    0.520 &   -0.383 &    0.000 &    0.490 &   -0.380 &   -0.000 &    0.455 &   -0.422 \\
		&   -0.000 &    0.694 &   -0.036 &   -0.000 &    0.666 &   -0.034 &   -0.000 &    0.679 &   -0.034 \\
		\hline
		Laug-cc-pVTZ &   -0.568 &   -0.000 &   -0.000 &   -8.247 &    0.000 &    0.000 &   -0.558 &    0.000 &    0.000 \\
		&   -0.000 &    0.501 &    1.912 &    0.000 &   -8.303 &    1.913 &   -0.000 &    0.493 &   -0.429 \\
		&   -0.000 &    0.699 &   -0.314 &    0.000 &   -8.813 &   -0.314 &   -0.000 &    0.700 &   -0.034 \\
		\hline
		Laug-cc-pVQZ &   -0.568 &   -0.000 &   -0.000 &   -8.200 &   -0.000 &   -0.000 &   -0.566 &   -0.000 &   -0.000 \\
		&    0.000 &    0.501 &    1.878 &   -0.000 &   -8.406 &    1.880 &   -0.000 &    0.498 &   -0.422 \\
		&   -0.000 &    0.698 &   -0.314 &   -0.000 &   -8.861 &   -0.315 &   -0.000 &    0.697 &   -0.035 \\
		\hline 
	\end{tabular} 
	\caption{H$_2$O$_2$ at a dihedral angle of 120$^\circ$: Basis set convergence of the cartesian anapole susceptibility tensor computed with GHF. L = London atomic orbitals. Contracted basis sets have been used. $\mathcal{M}$, $\mathcal{M'}$ and $\mathcal{M''}$ are defined in Eqs.~\eqref{Msus_def}, \eqref{Mprimesus_def} and \eqref{Mdblprimesus_def} respectively.}
	\label{H2O2_GHF_M_con}
\end{table}

\begin{table}
	\renewcommand{\arraystretch}{0.7}
	\begin{tabular}{|c|c|c|c|c|c|c|c|c|c|}
		\hline 
		Basis			& \multicolumn{3}{c|}{$\mathcal{M}$} &\multicolumn{3}{c|}{$\mathcal{M'}$} & \multicolumn{3}{c|}{$\mathcal{M''}$} \\
		\hline
		uSTO-3G &    0.570 &   -0.000 &   -0.000 &    1.463 &   -0.000 &    0.000 &    0.571 &   -0.000 &    0.000 \\
		&   -0.000 &   -0.698 &   -7.063 &   -0.000 &   25.805 &   -7.063 &   -0.000 &   -0.699 &   -1.348 \\
		&   -0.000 &    1.629 &    0.857 &   -0.000 &   32.946 &    0.857 &   -0.000 &    1.628 &   -0.018 \\
		\hline
		ucc-pVDZ &    0.019 &   -0.000 &   -0.000 &    0.048 &    0.000 &    0.000 &    0.021 &   -0.000 &   -0.000 \\
		&   -0.000 &   -0.078 &   -0.711 &   -0.000 &   -0.109 &   -0.711 &   -0.000 &   -0.080 &   -0.711 \\
		&   -0.000 &    0.935 &   -0.018 &   -0.000 &    0.927 &   -0.018 &   -0.000 &    0.934 &   -0.018 \\
		\hline
		ucc-pVTZ &   -0.371 &   -0.000 &   -0.000 &   -0.342 &   -0.000 &   -0.000 &   -0.370 &   -0.000 &    0.000 \\
		&   -0.000 &    0.314 &   -0.539 &    0.000 &    0.282 &   -0.539 &   -0.000 &    0.312 &   -0.539 \\
		&   -0.000 &    0.758 &   -0.026 &   -0.000 &    0.750 &   -0.026 &   -0.000 &    0.757 &   -0.026 \\
		\hline
		uaug-cc-pVDZ &   -0.451 &   -0.000 &   -0.000 &   -0.425 &    0.000 &    0.000 &   -0.450 &    0.000 &    0.000 \\
		&   -0.000 &    0.395 &   -0.466 &    0.000 &    0.366 &   -0.465 &   -0.000 &    0.394 &   -0.465 \\
		&   -0.000 &    0.724 &   -0.034 &   -0.000 &    0.717 &   -0.034 &   -0.000 &    0.724 &   -0.034 \\
		\hline
		uaug-cc-pVTZ &   -0.559 &   -0.000 &   -0.000 &   -0.530 &   -0.000 &   -0.000 &   -0.558 &   -0.000 &   -0.000 \\
		&   -0.000 &    0.497 &   -0.439 &    0.000 &    0.464 &   -0.439 &    0.000 &    0.495 &   -0.439 \\
		&   -0.000 &    0.707 &   -0.034 &    0.000 &    0.699 &   -0.034 &    0.000 &    0.706 &   -0.034 \\
		\hline
		uaug-cc-pVQZ &   -0.567 &   -0.000 &   -0.000 &   -0.537 &    0.000 &    0.000 &   -0.566 &   -0.000 &   -0.000 \\
		&   -0.000 &    0.500 &   -0.424 &   -0.000 &    0.467 &   -0.424 &   -0.000 &    0.499 &   -0.424 \\
		&   -0.000 &    0.698 &   -0.035 &   -0.000 &    0.690 &   -0.035 &   -0.000 &    0.698 &   -0.035 \\
		\hline
		LuSTO-3G &   -0.589 &   -0.000 &   -0.000 &    0.080 &   -0.000 &    0.000 &   -0.221 &    0.000 &   -0.000 \\
		&   -0.000 &    0.505 &   -4.917 &   -0.000 &   27.161 &   -5.974 &   -0.000 &    0.050 &   -0.902 \\
		&   -0.000 &    0.576 &    1.910 &    0.000 &   31.585 &    0.772 &   -0.000 &    1.105 &   -0.033 \\
		\hline
		Lucc-pVDZ &   -0.553 &   -0.000 &   -0.000 &   -0.524 &    0.000 &    0.000 &   -0.297 &   -0.000 &   -0.000 \\
		&   -0.000 &    0.509 &   -0.339 &    0.000 &    0.475 &   -0.348 &   -0.000 &    0.225 &   -0.598 \\
		&   -0.000 &    0.617 &   -0.060 &   -0.000 &    0.568 &   -0.045 &   -0.000 &    0.790 &   -0.026 \\
		\hline
		Lucc-pVTZ &   -0.557 &   -0.000 &   -0.000 &   -0.534 &    0.000 &    0.000 &   -0.450 &    0.000 &    0.000 \\
		&   -0.000 &    0.505 &   -0.397 &   -0.000 &    0.478 &   -0.402 &    0.000 &    0.384 &   -0.510 \\
		&   -0.000 &    0.655 &   -0.047 &    0.000 &    0.636 &   -0.041 &    0.000 &    0.725 &   -0.029 \\
		\hline
		Luaug-cc-pVDZ &   -0.580 &   -0.000 &   -0.000 &   -0.548 &   -0.000 &   -0.000 &   -0.508 &   -0.000 &    0.000 \\
		&   -0.000 &    0.513 &   -0.403 &   -0.000 &    0.483 &   -0.398 &   -0.000 &    0.449 &   -0.434 \\
		&   -0.000 &    0.698 &   -0.036 &    0.000 &    0.679 &   -0.033 &   -0.000 &    0.689 &   -0.034 \\
		\hline
		Luaug-cc-pVTZ &   -0.570 &   -0.000 &   -0.000 &   -0.540 &    0.000 &    0.000 &   -0.561 &    0.000 &   -0.000 \\
		&   -0.000 &    0.503 &   -0.418 &   -0.000 &    0.469 &   -0.417 &   -0.000 &    0.495 &   -0.432 \\
		&   -0.000 &    0.703 &   -0.036 &    0.000 &    0.695 &   -0.036 &   -0.000 &    0.702 &   -0.034 \\
		\hline
		Luaug-cc-pVQZ &   -0.567 &   -0.000 &   -0.000 &   -0.537 &   -0.000 &   -0.000 &   -0.566 &   -0.000 &    0.000 \\
		&   -0.000 &    0.501 &   -0.419 &    0.000 &    0.467 &   -0.419 &   -0.000 &    0.498 &   -0.422 \\
		&   -0.000 &    0.699 &   -0.035 &   -0.000 &    0.691 &   -0.035 &   -0.000 &    0.697 &   -0.035 \\
		\hline 
	\end{tabular} 
	\caption{H$_2$O$_2$ at a dihedral angle of 120$^\circ$: Basis set convergence of the cartesian anapole susceptibility tensor computed with GHF. L = London atomic orbitals. Uncontracted basis sets have been used. $\mathcal{M}$, $\mathcal{M'}$ and $\mathcal{M''}$ are defined in Eqs.~\eqref{Msus_def}, \eqref{Mprimesus_def} and \eqref{Mdblprimesus_def} respectively.}
	\label{H2O2_GHF_M_uncon}
\end{table}

\subsection{CHFXY}

The size effects of the orbital anapole moments and their relation to chirality have been studied earlier~\cite{Tellgren2013} for a series of halomethanes, CHFXY with X,Y = Cl, Br.
We have carried out computations including the spin effects on the same sample set.
Here too, we have used uncontracted normalized basis sets.
In Tables~\ref{CHFCl2_GHF_A_uncon}-\ref{CHFBr2_GHF_A_uncon}, the values for $\mathcal{A}$ and $\mathcal{A}'$ are reported.
For $\mathcal{A}$, we only calculate the diagonal elements of the tensor to avoid computing too many finite-field points.
The values of $\mathcal{A}$ and $\mathcal{A}'$ increase in magnitude from CHFCl$_2$ to CHFClBr to CHFBr$_2$ indicating significant size effects.
With increasing size, the discrepancy between $\mathcal{A}$ and $\mathcal{A}'$ also increases.
The CHFXY molecule is placed in a coordinate system such that C lies at the origin, H on the z-axis and F in the yz-plane. The exact geometries are reported in the Supplementary Information.
With this orientation, the off-diagonal elements $\mathcal{A}_{xy}'$, $\mathcal{A}_{yx}'$, $\mathcal{A}_{xz}'$, and $\mathcal{A}_{zx}'$, but not $\mathcal{A}'_{yz}$, are zero in CHFCl$_2$ and CHFBr$_2$ due to symmetry. By contrast, all elements of the $\mathcal{A}'$ tensor are seen in the chiral CHFClBr.

The basis set convergence is very poor with ordinary basis sets with values in the largest basis being off by factors of 5-8 from the best estimate with LAOs in the same basis.
The deviation between $\mathcal{A}$ and $\mathcal{A}'$ when using LAOs is found to be larger for GHF than for RHF.

\begin{table}
	\renewcommand{\arraystretch}{0.7}
\begin{tabular}{|c|c|c|c|c|c|c|}
	\hline 
	Basis			& \multicolumn{3}{c|}{$\mathcal{A}$} & \multicolumn{3}{c|}{$\mathcal{A'}$} \\
	\hline
	uSTO-3G & -118.429 & N/A & N/A & -118.425 &    0.000 &   -0.000 \\
	& N/A & -215.102 & N/A &    0.000 & -215.090 &   29.875 \\
	& N/A & N/A & -259.845 &   -0.000 &   29.875 & -259.829 \\
	\hline
	ucc-pVDZ &  -80.650 & N/A & N/A &  -80.646 &    0.000 &   -0.000 \\
	& N/A & -155.533 & N/A &    0.000 & -155.520 &   23.669 \\
	& N/A & N/A & -184.587 &    0.000 &   23.670 & -184.570 \\
	\hline
	ucc-pVTZ &  -48.206 & N/A & N/A &  -48.204 &   -0.000 &    0.000 \\
	& N/A & -105.138 & N/A &    0.000 & -105.126 &   18.053 \\
	& N/A & N/A & -125.091 &    0.000 &   18.054 & -125.077 \\
	\hline
	uaug-cc-pVDZ &  -41.734 & N/A & N/A &  -41.738 &   -0.000 &   -0.000 \\
	& N/A &  -99.461 & N/A &    0.000 &  -99.465 &   18.199 \\
	& N/A & N/A & -118.646 &   -0.000 &   18.199 & -118.652 \\
	\hline
	uaug-cc-pVTZ &  -32.208 & N/A & N/A &  -32.209 &   -0.000 &   -0.000 \\
	& N/A &  -83.467 & N/A &   -0.000 &  -83.463 &   16.191 \\
	& N/A & N/A & -100.006 &   -0.000 &   16.191 & -100.003 \\
	\hline
	uaug-cc-pVQZ &  -22.126 & N/A & N/A &  -22.126 &   -0.000 &   -0.000 \\
	& N/A &  -59.965 & N/A &   -0.000 &  -59.961 &   12.050 \\
	& N/A & N/A &  -73.027 &   -0.000 &   12.051 &  -73.022 \\
	\hline
	LuSTO-3G &  -27.682 & N/A & N/A &  -37.774 &   -0.000 &    0.000 \\
	& N/A &  -38.525 & N/A &   -0.000 &  -57.434 &    6.283 \\
	& N/A & N/A &  -43.363 &    0.000 &    6.160 &  -66.115 \\
	\hline
	Lucc-pVDZ &  -23.693 & N/A & N/A &  -29.615 &    0.000 &   -0.000 \\
	& N/A &  -35.690 & N/A &   -0.000 &  -46.423 &    5.382 \\
	& N/A & N/A &  -41.051 &   -0.000 &    5.357 &  -54.045 \\
	\hline
	Lucc-pVTZ &  -13.458 & N/A & N/A &  -15.863 &    0.000 &    0.000 \\
	& N/A &  -26.123 & N/A &   -0.000 &  -30.249 &    4.747 \\
	& N/A & N/A &  -32.469 &    0.000 &    4.709 &  -37.545 \\
	\hline
	Luaug-cc-pVDZ &   -9.397 & N/A & N/A &  -10.098 &    0.000 &   -0.000 \\
	& N/A &  -22.389 & N/A &    0.000 &  -23.810 &    4.532 \\
	& N/A & N/A &  -29.357 &    0.000 &    4.491 &  -30.910 \\
	\hline
	Luaug-cc-pVTZ &   -6.122 & N/A & N/A &   -6.301 &   -0.000 &    0.000 \\
	& N/A &  -19.019 & N/A &   -0.000 &  -19.408 &    4.358 \\
	& N/A & N/A &  -25.907 &    0.000 &    4.357 &  -26.336 \\
	\hline
	Luaug-cc-pVQZ &   -5.325 & N/A & N/A &   -5.352 &   -0.000 &    0.000 \\
	& N/A &  -18.163 & N/A &   -0.000 &  -18.228 &    4.297 \\
	& N/A & N/A &  -25.006 &   -0.000 &    4.300 &  -25.072 \\
	\hline
\end{tabular} 
\caption{CHFCl$_2$ : Basis set convergence of the cartesian anapole susceptibility tensor computed with GHF. L = London atomic orbitals, u = uncontracted. $\mathcal{A}$ and $\mathcal{A'}$ are defined in Eqs.~\eqref{Asus_def} and \eqref{Aprimesus_def} respectively.}
\label{CHFCl2_GHF_A_uncon}
\end{table}

\begin{table}
	\renewcommand{\arraystretch}{0.8}
	\begin{tabular}{|c|c|c|c|c|c|c|}
	\hline 
	Basis			& \multicolumn{3}{c|}{$\mathcal{A}$} & \multicolumn{3}{c|}{$\mathcal{A'}$} \\
	\hline
	uSTO-3G & -170.888 & N/A & N/A & -170.883 &   62.915 &   39.246 \\
	& N/A & -349.579 & N/A &   62.922 & -349.556 &   51.454 \\
	& N/A & N/A & -415.077 &   39.252 &   51.455 & -415.046 \\
	\hline
	ucc-pVDZ & -118.315 & N/A & N/A & -118.311 &   46.888 &   29.446 \\
	& N/A & -254.354 & N/A &   46.892 & -254.334 &   39.752 \\
	& N/A & N/A & -298.734 &   29.450 &   39.753 & -298.709 \\
	\hline
	ucc-pVTZ &  -81.786 & N/A & N/A &  -81.782 &   43.718 &   27.740 \\
	& N/A & -195.693 & N/A &   43.722 & -195.675 &   33.162 \\
	& N/A & N/A & -229.790 &   27.743 &   33.163 & -229.769 \\
	\hline
	uaug-cc-pVDZ &  -76.973 & N/A & N/A &  -76.978 &   46.504 &   29.406 \\
	& N/A & -195.320 & N/A &   46.505 & -195.326 &   34.409 \\
	& N/A & N/A & -229.682 &   29.407 &   34.410 & -229.690 \\
	\hline
	uaug-cc-pVTZ &  -63.496 & N/A & N/A &  -63.498 &   42.057 &   26.681 \\
	& N/A & -169.600 & N/A &   42.059 & -169.599 &   30.851 \\
	& N/A & N/A & -199.893 &   26.683 &   30.853 & -199.891 \\
	\hline
	LuSTO-3G &  -34.853 & N/A & N/A &  -47.504 &    6.371 &    4.027 \\
	& N/A &  -50.394 & N/A &    6.322 &  -75.982 &    8.767 \\
	& N/A & N/A &  -56.250 &    3.957 &    8.608 &  -86.705 \\
	\hline
	Lucc-pVDZ &  -29.066 & N/A & N/A &  -36.282 &    5.528 &    3.615 \\
	& N/A &  -46.331 & N/A &    5.572 &  -60.852 &    7.528 \\
	& N/A & N/A &  -52.893 &    3.613 &    7.491 &  -70.224 \\
	\hline
	Lucc-pVTZ &  -17.274 & N/A & N/A &  -20.656 &    5.472 &    3.636 \\
	& N/A &  -35.672 & N/A &    5.522 &  -42.639 &    6.865 \\
	& N/A & N/A &  -43.370 &    3.672 &    6.833 &  -51.686 \\
	\hline
	Luaug-cc-pVDZ &  -12.531 & N/A & N/A &  -14.441 &    5.745 &    3.695 \\
	& N/A &  -31.264 & N/A &    5.756 &  -35.986 &    6.783 \\
	& N/A & N/A &  -39.691 &    3.725 &    6.738 &  -45.029 \\
	\hline
	Luaug-cc-pVTZ &   -8.706 & N/A & N/A &   -9.739 &    5.313 &    3.467 \\
	& N/A &  -27.440 & N/A &    5.321 &  -30.135 &    6.440 \\
	& N/A & N/A &  -35.766 &    3.470 &    6.437 &  -38.853 \\
	\hline
\end{tabular}
\caption{CHFClBr : Basis set convergence of the cartesian anapole susceptibility tensor computed with GHF. L = London atomic orbitals, u = uncontracted. $\mathcal{A}$ and $\mathcal{A'}$ are defined in Eqs.~\eqref{Asus_def} and \eqref{Aprimesus_def} respectively.}
\label{CHFClBr_GHF_A_uncon}
\end{table}

\begin{table}
	\renewcommand{\arraystretch}{0.8}
	\begin{tabular}{|c|c|c|c|c|c|c|}
		\hline 
		Basis			& \multicolumn{3}{c|}{$\mathcal{A}$} & \multicolumn{3}{c|}{$\mathcal{A'}$} \\
		\hline
		uSTO-3G & -226.336 & N/A & N/A & -226.330 &    0.000 &    0.000 \\
		& N/A & -477.912 & N/A &   -0.000 & -477.882 &   74.504 \\
		& N/A & N/A & -565.218 &   -0.000 &   74.505 & -565.177 \\
		\hline
		ucc-pVDZ & -157.930 & N/A & N/A & -157.925 &   -0.000 &    0.000 \\
		& N/A & -348.371 & N/A &   -0.000 & -348.346 &   56.869 \\
		& N/A & N/A & -408.674 &    0.000 &   56.871 & -408.641 \\
		\hline
		ucc-pVTZ & -117.270 & N/A & N/A & -117.268 &   -0.000 &   -0.000 \\
		& N/A & -281.856 & N/A &    0.000 & -281.835 &   49.186 \\
		& N/A & N/A & -330.595 &    0.000 &   49.187 & -330.571 \\
		\hline
		uaug-cc-pVDZ & -114.366 & N/A & N/A & -114.375 &   -0.000 &    0.000 \\
		& N/A & -286.525 & N/A &   -0.000 & -286.537 &   51.683 \\
		& N/A & N/A & -336.685 &   -0.000 &   51.684 & -336.699 \\
		\hline
		uaug-cc-pVTZ &  -96.688 & N/A & N/A &  -96.694 &   -0.000 &    0.000 \\
		& N/A & -251.541 & N/A &   -0.000 & -251.542 &   46.518 \\
		& N/A & N/A & -296.165 &   -0.000 &   46.520 & -296.165 \\
		\hline
		LuSTO-3G &  -42.222 & N/A & N/A &  -57.647 &   -0.000 &   -0.000 \\
		& N/A &  -62.485 & N/A &    0.000 &  -94.537 &   11.518 \\
		& N/A & N/A &  -69.355 &    0.000 &   11.314 & -107.409 \\
		\hline
		Lucc-pVDZ &  -34.493 & N/A & N/A &  -43.084 &    0.000 &   -0.000 \\
		& N/A &  -56.850 & N/A &   -0.000 &  -75.027 &    9.896 \\
		& N/A & N/A &  -64.565 &   -0.000 &    9.848 &  -86.126 \\
		\hline
		Lucc-pVTZ &  -21.162 & N/A & N/A &  -25.602 &    0.000 &   -0.000 \\
		& N/A &  -45.142 & N/A &   -0.000 &  -54.862 &    9.169 \\
		& N/A & N/A &  -54.149 &   -0.000 &    9.142 &  -65.609 \\
		\hline
		Luaug-cc-pVDZ &  -15.818 & N/A & N/A &  -19.042 &   -0.000 &   -0.000 \\
		& N/A &  -40.094 & N/A &   -0.000 &  -47.984 &    9.254 \\
		& N/A & N/A &  -49.923 &   -0.000 &    9.203 &  -58.952 \\
		\hline
		Luaug-cc-pVTZ &  -11.435 & N/A & N/A &  -13.393 &   -0.000 &   -0.000 \\
		& N/A &  -35.816 & N/A &   -0.000 &  -40.723 &    8.732 \\
		& N/A & N/A &  -45.533 &   -0.000 &    8.730 &  -51.213 \\	
		\hline
	\end{tabular} 
\caption{CHFBr$_2$ : Basis set convergence of the cartesian anapole susceptibility tensor computed with GHF. L = London atomic orbitals, u = uncontracted. $\mathcal{A}$ and $\mathcal{A'}$ are defined in Eqs.~\eqref{Asus_def} and \eqref{Aprimesus_def} respectively.}
\label{CHFBr2_GHF_A_uncon}
\end{table}

\section{Summary and Conclusion} \label{summary}

We have reported a non-perturbative GHF implementation for molecules subject to finite, non-uniform magnetic fields. The implementation has been applied to study joint orbital and spin effects on energies and anapole susceptibilities of several molecules. The anapole susceptibilities provide a convenient quantification of the sensitivity to transverse magnetic field gradients.

By comparing GHF and RHF/UHF results, we are able to evaluate the relative importance of spin and orbital effects. Spin symmetry breaking due to magnetic field gradients has also been directly illustrated using the spin quantum number in H$_2$. In general, spin effects on the anapole susceptibility are large and have a consistent direction.
We have shown on theoretical grounds that spin effects always lower the second-order energy, at least for molecules that are closed shell singlets in the absence of magnetic fields. For molecules such as those in the present work, with generalized orbital diamagnetism, the orbital and spin effects on the tensor $\mathcal{A}$ must therefore oppose each other. An interesting pattern in the numerical results is that the difference between GHF and RHF, when quadratic in the transverse field gradient, is equal to \emph{half} the spin-Zeeman energy. A theoretical explanation of this fact has been derived in Sec.~\ref{SpinandOrb}.

Moreover, as has been shown previously for the orbital effects in isolation, the use of London atomic orbitals dramatically accelerates basis set convergence. This remains true for the spin effects. In addition, our results indicate that decontraction of the basis sets substantially increases accuracy.

\section*{Acknowledgements}

This work was supported by the Research Council of Norway through Grant No. 240674 and CoE Hylleraas Centre for Molecular Sciences Grant No. 262695, and the European  Union’s Horizon 2020 research and innovation programme under the Marie Sk{\l}odowska-Curie grant agreement No.~745336. This work has also received support from the Norwegian Supercomputing Program (NOTUR) through a grant of computer time (Grant No.~NN4654K).

\newpage

\end{document}